\documentclass[
    reprint,
    amsmath,amssymb,superscriptaddress,
    aps,
    pre,
]{revtex4-2}

\usepackage{bm}
\usepackage{braket}
\usepackage{mathtools}
\usepackage[hidelinks]{hyperref}
\usepackage{xcolor}

\newcommand{\Arg}{(\bm{\mu})}
\newcommand{\ArgI}[1]{(\bm{\mu}_{#1})}

\newcommand{\rb}{_\mathrm{rb}}
\newcommand{\cut}{\mathtt{cut}}

\DeclareMathOperator{\vspan}{span}
\DeclareMathOperator*{\argmax}{arg\,max}
\DeclareMathOperator*{\argmin}{arg\,min}

\begin{document}

% TITLE
\title{Reduced basis surrogates for quantum spin systems based on tensor networks}
\date{\today}
% \the\textwidth % write textwidth to document for scale-free figures

% AUTHORS
\author{Paul Brehmer}
\email{paul.brehmer@rwth-aachen.de}
\affiliation{Institute for Theoretical Solid State Physics, RWTH Aachen University, Otto-Blumenthal-Str.~26, 52074 Aachen, Germany}

\author{Michael F. Herbst}
\email{michael.herbst@epfl.ch}
\affiliation{Mathematics for Materials Modelling, Institute of Mathematics \& Institute of Materials, EPFL, CH-1015 Lausanne}

\author{Stefan Wessel}
\email{wessel@physik.rwth-aachen.de}
\affiliation{Institute for Theoretical Solid State Physics, RWTH Aachen University, Otto-Blumenthal-Str.~26, 52074 Aachen, Germany}

\author{Matteo Rizzi}
\email{m.rizzi@fz-juelich.de}
\affiliation{Forschungszentrum J\"ulich GmbH, Institute of Quantum Control, Peter Gr\"unberg Institut (PGI-8), 52425 J\"ulich, Germany}
\affiliation{Institute for Theoretical Physics, University of Cologne, D-50937 K\"oln, Germany}

\author{Benjamin Stamm}
\email{best@ians.uni-stuttgart.de}
\affiliation{Institute of Applied Analysis and Numerical Simulation, University of Stuttgart, 70569 Stuttgart, Germany}

% ABSTRACT
\begin{abstract}
    Within the reduced basis methods approach, an effective low-dimensional subspace of a quantum many-body Hilbert space is constructed in order to investigate, e.g., the ground-state phase diagram.
    The basis of this subspace is built from solutions of snapshots, i.e., ground states corresponding to particular and well-chosen parameter values.
    Here, we show how a greedy strategy to assemble the reduced basis and thus to select the parameter points can be implemented based on matrix-product-state (MPS) calculations.
    Once the reduced basis has been obtained, observables required for the computation of phase diagrams can be computed with a computational complexity independent of the underlying Hilbert space for any parameter value.
    We illustrate the efficiency and accuracy of this approach for different one-dimensional quantum spin-1 models, including anisotropic as well as biquadratic exchange interactions, leading to rich quantum phase diagrams. 
\end{abstract}

% MAKE TITLE
\maketitle

% MAIN BODY
\section{Introduction}

A central topic in modern condensed matter theory is the exploration of ground-state phase diagrams of quantum many-body systems.
They uncover the rich collective behavior of various physical models and often harbor interesting states of matter that emerge from interaction effects. 
However, only a small number of many-body Hamiltonians are solvable analytically, and in most cases one needs to resort to computational methods to explore strongly correlated quantum systems. 
Moreover, in order to relate microscopic models to experimental findings, it is often required to extend beyond the most basic model Hamiltonians that describe an emerging quantum many-body effect by taking into account additional interaction terms or anisotropy effects.
An illustrative example from quantum magnetism is provided by its most basic model system --- the one-dimensional  spin chain.
In particular, it is by now well established, dating back to the seminal works by Haldane~\cite{Haldane1981,Haldane1983,Haldane1983B}, that the spin-1 Heisenberg chain exhibits a gapped, quantum-disordered ground state with symmetry-protected topological order~\cite{Gu2009,Pollmann2012}.
A quantitative description of typical Haldane 
spin chain materials requires to  account for both uniaxial and rhombic-type single-ion anisotropies, yielding a rich overall ground-state phase diagram~\cite{tzengQuantumPhaseTransitions2017,Ren2018}.

In the numerical study of complex phase diagrams of quantum many-body systems we then face an exacerbated computational challenge: 
Not only is solving the Hamiltonian already demanding due the curse of dimensionality, but the solutions have to be obtained on many parameter points.
Typically, this leads either to the compromise of resorting to low-resolution scans of the phase diagram or to the brute-force way of solving the problem in massively parallel computations.
When these approaches become impractical, we advocate here that so-called reduced basis (RB) methods~\cite{hesthavenCertifiedReducedBasis2016,quarteroniReducedBasisMethods2016} provide a powerful third option, as we started exploring recently~\cite{herbstSurrogateModelsQuantum2022}.
Harnessing the linear dependence of eigenstates across the phase diagram, a low-dimensional surrogate model is constructed based on a few select ground states at different parameter points.
Using the surrogate, one is then able to evaluate observables independent of the Hilbert space dimension on any point in the phase diagram, providing accuracy and efficiency at the same time.

The RB approach for parametrized eigenvalue problems originated from the structural analysis of mechanical systems in the late nineties~\cite{aktasReducedBasisEigenvalue1998,nairImprovedFirstOrderApproximation1998}.
It has since then migrated to the applied mathematics community with the pioneering work presented in Ref.~\cite{machielsOutputBoundsReducedbasis2000}. 
While the RB method became popular for studying parametrized partial differential equations (PDEs)~\cite{hesthavenCertifiedReducedBasis2016,quarteroniReducedBasisMethods2016}, with many engineering applications and maturity for industrial applications, the RB method for parametrized eigenvalue problems had not received much attention thus far.
Only years later, the RB methodology developed in Ref.~\cite{machielsOutputBoundsReducedbasis2000} was further extended and generalized to account for better error control and more general settings~\cite{fumagalliReducedBasisApproximation2016,horgerSimultaneousReducedBasis2017,plaReducedBasisMethod2019,Pichi2020}.

In the context of quantum many-body physics, RB methods for eigenvalue problems found application only recently and are most prominent in the nuclear physics community~\cite{Ekstroem2019,Demol2020,Koenig2020,Furnstahl2020,Drischler2021,Bonilla2022}, where their use emerged with the eigenvector continuation (EC) method~\cite{frameEigenvectorContinuationSubspace2018, Sarkar2021, sarkarSelflearningEmulatorsEigenvector2022}. 
EC was retrospectively identified as belonging to the family of RB methods, and more broadly, to the field of model order reduction~\cite{Melendez2022, Bonilla2022}.
In further physical applications, RB surrogate models, referred to as \textit{emulators} in EC language, proved viable in determining phase diagrams of quantum spin systems~\cite{herbstSurrogateModelsQuantum2022}.
Furthermore, EC emulators were used as a subspace diagonalization method on quantum computers~\cite{francisSubspaceDiagonalizationQuantum2022}, for emulating superconducting phenomena~\cite{baranReducedBasisEmulation2022, CompanysFranzke2023}, as well as in quantum chemistry applications~\cite{MejutoZaera2023}.

In this article, we expand upon our previous work~\cite{herbstSurrogateModelsQuantum2022} that utilizes exact diagonalization (ED) techniques in combination with a greedy RB approach.
As suggested therein, we apply the density matrix renormalization group (DMRG)~\cite{White1992,White1993} to treat spin chain Hamiltonians and perform all necessary vector operations using MPSs~\cite{Fannes1992,Kluemper1993,Ostlund1995,schollwoeckDensitymatrixRenormalizationGroup2011}, which belong to the family of tensor network states~\cite{Verstraete2008,orusPracticalIntroductionTensor2014} that our approach could be naturally extended to (at least for finite system sizes).
While this approach has been very recently adopted in~\cite{baranReducedBasisEmulation2022}, we focus on methodological aspects and develop the combined RB-MPS method in application to quantum spin systems.
In using compressed MPSs, we gain access to larger many-body systems, allowing us to probe whether RB methods remain a viable approach in the light of larger Hilbert spaces and more complex parameter domains. We made a related open-source software package available~\cite{GitHubRepo}.

The remainder of this paper is organized as follows:
In Sec.~\ref{sec:method} we review the greedy RB algorithm and introduce notation, before explaining the combined RB-MPS approach.
This allows us to apply the method to different spin-1 chain Hamiltonians in Sec.~\ref{sec:results}, namely the Haldane chain with uniaxial and rhombic single-ion anisotropies as well as the bilinear-biquadratic model with a uniaxial anisotropy, where we scan the ground-state phase diagrams for various correlations.
Section~\ref{sec:technical-discussion} discusses the accuracy of the surrogate models in different settings as well as the effects of MPS approximations on the RB framework, after which we conclude in Sec.~\ref{sec:conclusions}.

\section{Method} \label{sec:method}
\subsection{The RB framework}
\paragraph*{Problem setting.}
We begin by defining the physical problems that we aim to treat in the RB approach.
To that end, we consider a generic stationary quantum many-body problem
\begin{equation} \label{eqn:schroedinger}
    H\Arg \ket{\bm\Psi\Arg} = E\Arg \ket{\bm\Psi\Arg},
\end{equation}
that consists of finding the ground-state energy $E\Arg$ and its corresponding $m$ ground states $\ket{\bm\Psi\Arg} = (\ket{\Psi^{[1]}\Arg}, \dots, \ket{\Psi^{[m]}\Arg})$.
The many-body Hilbert space $\mathcal{H} = \mathbb{C}^\mathcal{N}$ under consideration is assumed finite but high-dimensional since $\mathcal{N} \gg 1$ grows exponentially with the number of physical constituents.
We assume the eigenvalue problem to be parametrized by a vector of physical model parameters $\bm\mu \in \mathbb{P}$ that resides in the parameter space $\mathbb{P}$, i.e., for each $\bm\mu$ we obtain a new Hamiltonian to solve.
In order to apply the RB framework, we consider a particular class of Hamiltonians that can be expressed as \textit{affine decompositions}
\begin{equation}
    H\Arg = \sum_{q=1}^Q \theta_q\Arg\, H_q,
\end{equation}
with a number of terms $Q$ independent of the Hilbert space dimension $\mathcal{N}$.
Such linear combinations distinguish between parameter-dependent coefficients $\theta_q : \mathbb{P} \to \mathbb{R}$ and parameter-independent Hermitian matrices $H_q : \mathcal{H} \to \mathcal{H}$.
Not only the Hamiltonian must be affinely decomposable, but in fact all observables that we may want to measure are assumed to be of the form
\begin{equation} \label{eqn:observable}
    O(\bm\mu; p) = \sum_{r=1}^R \alpha_r(\bm\mu; p)\, O_r,
\end{equation}
in terms of operators $O_r : \mathcal{H} \to \mathcal{H}$, and where the coefficients $\alpha_r(\bm\mu; p) \in \mathbb{C}$ now include additional parameters $p$ that are separate from $\mathbb{P}$.
This generalization is required, e.g., when studying observables in Fourier space.
Note that the individual operators $O_r$ need not necessarily be Hermitian.
To then extract physical information, one measures affine decompositions by taking the expectation value
\begin{equation}
    \braket{O(\bm\mu; p)} = \sum_{r=1}^R \alpha_r(\bm\mu; p)\,
    \frac{1}{m} \sum_{i=1}^m \braket{\Psi^{[i]}\Arg | O_r | \Psi^{[i]}\Arg},
\end{equation}
where we average over the degenerate ground-state subspace, if needed.

In the particular setting we consider here, the goal is to scan domains of $\mathbb{P}$ with fine resolution, i.e., to solve the many-body problem in Eq.~\eqref{eqn:schroedinger} on a large set of parameter points and then, based on the solutions, measure observables on the same parameter domain.
This is addressed by the RB approach in two steps:
In the first step, which is referred to as the \textit{offline stage} in RB parlance, a surrogate model is assembled based on a small number of ground states across the parameter space, whereas in the second step, the so-called \textit{online stage}, observable measurements are efficiently obtained.
This approach is warranted by the insight that ground states at different parameter points tend to show significant linear dependence.
While being demonstrated in various numerical applications (see e.g.~\cite{herbstSurrogateModelsQuantum2022,Bonilla2022}), this can also be reasoned in terms of analytical continuation in the context of eigenvector continuation~\cite{frameEigenvectorContinuationSubspace2018,Sarkar2021}.
One particular strategy to build RB surrogate models features a greedy mechanism~\cite{herbstSurrogateModelsQuantum2022,quarteroniReducedBasisMethods2016,hesthavenCertifiedReducedBasis2016} to sample the parameter domain, which we want to review next.

\paragraph*{Offline stage.}
The goal in the offline stage is to construct a low-dimensional \textit{reduced basis space}
\begin{equation}
    \mathbb{V}_n \coloneqq \vspan \big\{\!\ket{\bm\Psi\ArgI{1}}, \dots, \ket{\bm{\Psi}\ArgI{n}}\!\big\}
\end{equation}
that is spanned by degenerate subspaces extracted at $n$ different parameter points $\{\bm\mu_1, \dots, \bm\mu_n\}$.
Here we refer to $\ket{\bm\Psi\ArgI{j}} = (\ket{\Psi^{[1]}\ArgI{j}},\dots,\ket{\Psi^{[m_j]}\ArgI{j}})$ as a ``snapshot'', such that one snapshot contains $m_j \geq 1$ states.
Since the snapshots might share linear dependent modes, the reduced basis dimension $N \coloneqq \dim \mathbb{V}_n \leq M \coloneqq \sum_{j=1}^n m_j$ can in general be smaller than the total number of obtained ground states $M$.
For the single ground-state solutions that span the RB space we use the shorthand $\ket{\Psi_j}$ where $j=1,\dots,N$.
Moreover, the solver used to obtain the lowest eigenvalue and eigenvectors of $H\Arg$ is called the ``truth solver'', which is to be understood in the sense that it can obtain the true solution of the Hamiltonian at any point in the considered parameter domain, up to high numerical accuracy.
Beyond that, we do not need to further specify the actual method yet, i.e., the RB approach is solver-agnostic.
The idea is then to represent ground states at any other parameter point $\bm\mu\in\mathbb{P}$ as a linear combination of snapshots
\begin{equation} \label{eqn:interpolated-ground-state}
    \ket{\Phi\rb\Arg} = \sum_{j=1}^{N} a_j\Arg \ket{\Psi_j},
\end{equation}
with parameter-dependent coefficients $a_j\Arg \in \mathbb{C}$.
This prompts the following two questions: First, how do we generate the RB space $\mathbb{V}_n$ and in particular how are the parameter points chosen, and second, how do we determine the coefficients $a_j\Arg$?

To answer the first question, we turn to the greedy algorithm.
We restrict the snapshots to a subset of all parameter points, which we call the \textit{training grid} $\Xi_\mathrm{train} \subset \mathbb{P}$, that preselects the domain on which the RB will be ``trained''.
Starting from a suitable first parameter point $\bm\mu_1$ and its corresponding snapshot $\ket{\bm\Psi\ArgI{1}}$, the RB is constructed in an inductive manner.
Namely at iteration $n$, using the current RB space $\mathbb{V}_n$, we select the next parameter point $\bm\mu_{n+1}$ where a truth solve will be performed.
Based on the ground states that are contained in the RB space
\begin{equation} \label{eqn:variational-ground-state}
    \ket{\bm\Phi\rb\Arg} = \argmin_{\ket{\Phi} \in \mathbb{V}_n}
    \frac{\braket{\Phi | H\Arg | \Phi}}{\braket{\Phi | \Phi}},\quad
    \forall \bm{\mu} \in \Xi_\mathrm{train},
\end{equation}
solved at all training points, we select the parameter point that meets the \textit{greedy condition}
\begin{equation}
    \bm{\mu}_{n+1} = \argmax_{\bm{\mu} \in \Xi_\mathrm{train}}\, \mathrm{Res}_n\Arg,
\end{equation}
i.e., which corresponds to the snapshot that maximizes the \textit{residual}
\begin{equation} \label{eqn:residual}
    \mathrm{Res}_n\Arg \coloneqq \sqrt{\sum_{i=1}^m \big\lVert H\Arg \ket{\Phi_\mathrm{rb}^{[i]}\Arg} -
    E\rb\Arg \ket{\Phi_\mathrm{rb}^{[i]}\Arg} \big\rVert^2 }.
\end{equation}
In this way, we sample first the parts of $\Xi_\mathrm{train}$ where the RB performs worst, meaning that the surrogate ground states and energies least accurately fulfill the eigenvalue problem of $H\Arg$.
Notice the dependency of $\ket{\bm\Phi\rb}$ and $E\rb$ on the number of greedy iterations since they were obtained from the $n$-snapshot RB space $\mathbb{V}_n$.
Following this, we perform a truth solve at $\bm\mu_{n+1}$ and obtain a snapshot $\ket{\bm\Psi\ArgI{n+1}}$ that is appended to the basis, forming $\mathbb{V}_{n+1}$.
This iteration is repeated until an adequate exit condition, e.g., a target residual accuracy, is reached.

In the current formulation, it seems difficult and expensive to solve Eq.~\eqref{eqn:variational-ground-state} in order to determine the greedy condition, i.e., to compute the residual on all training points.
To clarify this and to answer the question of determining the linear coefficients of $\ket{\bm\Phi\rb\Arg}$, we now map the problem onto the low-dimensional RB space via the \textit{reduced basis} $B\!: \mathbb{V}_n \to \mathcal{H}$ by way of the Rayleigh-Ritz method.
To that end, we define the RB coefficients $\bm{\varphi}\rb\Arg \in \mathbb{C}^{N \times m}$ by relating them to
\begin{equation} \label{eqn:rb-ground-state}
    \ket{\bm\Phi\rb\Arg} = B \bm{\varphi}\rb\Arg \in \mathbb{C}^{\mathcal{N} \times m},
\end{equation}
corresponding to the Ritz vector, as well as the normalization matrix $b = B^\dagger B$ and the reduced Hamiltonian
\begin{equation}
    h\Arg = \sum_{q=1}^Q \theta_q\Arg\, h_q,\quad
    h_q = B^\dagger H_q B.
\end{equation}
Then we can reformulate the variational problem 
of Eq.~\eqref{eqn:variational-ground-state} as a generalized eigenvalue problem
\begin{equation} \label{eqn:rb-evp}
    h\Arg\, \bm{\varphi}\rb\Arg = 
    E\rb\Arg\, b\, \bm{\varphi}\rb\Arg,
\end{equation}
solving only for the lowest eigenvalue and corresponding eigenvectors.
Indeed, by expressing the residual in terms of reduced quantities and using the eigenvalue Eq.~\eqref{eqn:rb-evp}, one finds the more efficient expression
\begin{align} \label{eqn:residual-cheap}
    \mathrm{Res}_n^2\Arg = &\sum_{q,q'=1}^Q \theta_q\Arg\, \theta_{q'}\Arg\,
    \sum_{i=1}^m \varphi_\mathrm{rb}^{[i]}\Arg^\dagger\, h_{qq'}\, \varphi_\mathrm{rb}^{[i]}\Arg\nonumber\\
    &- E^2\rb\Arg \sum_{i=1}^m \varphi_\mathrm{rb}^{[i]}\Arg^\dagger\, b\, \varphi_\mathrm{rb}^{[i]}\Arg,
\end{align}
where $h_{qq'} = B^\dagger H_q H_{q'} B$ are the reduced matrices of the $h^2(\bm\mu)$ affine decomposition.
We have thus obtained the greedy assembly algorithm, where computing the greedy condition boils down to the $N$-dimensional generalized eigenvalue problem of Eq.~\eqref{eqn:rb-evp} and all Hilbert space dimension dependent operations were isolated to the computation of $b$, $h_q$, $h_{qq'}$ and the truth solve, which are performed only once per iteration.

Finally, to determine $B$ itself, we might make the simple ansatz of using the snapshots as column vectors $[\ket{\bm\Psi\ArgI{1}} \cdots \ket{\bm\Psi\ArgI{n}}]$.
This, however, leads to a poorly-conditioned normalization $b$, due to the increasing linear dependence between the columns as we add snapshots, and quickly renders the solution of Eq.~\eqref{eqn:rb-evp} numerically unfeasible.
For that reason, one typically orthogonalizes the RB such that $b \simeq I$.
The orthogonalization procedure can be implemented on the level of the coefficients $\bm\varphi\rb$ or directly on the snapshots --- this depends on the truth solver and in particular the associated vector format, which we leave unspecified for now.

\paragraph*{Online stage.}
Once the offline stage is finished, we are left with the low-dimensional basis $B$ and the reduced quantities $b$, $h\Arg$ and $h^2\Arg$ as byproducts of the greedy assembly algorithm.
In order to compute expectation values of affine decompositions $O(\bm\mu; p)$, we compute the reduced matrices $o_r = B^\dagger O_r B$, so that we now operate entirely in RB space.
Namely, by taking the expectation value with the emulated ground state $\ket{\bm\Phi\rb\Arg}$, we obtain
\begin{equation} \label{eqn:rb-expectation}
    \braket{O(\bm{\mu}; p)}\rb = \sum_{r=1}^R \alpha_r(\bm{\mu}; p)\,
    \frac{1}{m} \sum_{i=1}^m \varphi_\mathrm{rb}^{[i]}\Arg^\dagger\, o_r\, \varphi^{[i]}_\mathrm{rb}\Arg.
\end{equation}
It becomes clear that we can, as advertised, evaluate the above on any parameter point $\bm\mu\in\mathbb{P}$, independent of $\mathcal{N}$ --- thanks to the affine decomposition, $o_r$ does not depend on $\bm\mu$ and we only need to evaluate the coefficient functions and compute the RB coefficients $\bm\varphi\rb\Arg$.
This again amounts to solving the $N$-dimensional generalized eigenvalue problem of Eq.~\eqref{eqn:rb-evp}.
For further details on the greedy RB approach, we refer to Refs.~\cite{herbstSurrogateModelsQuantum2022,quarteroniReducedBasisMethods2016,hesthavenCertifiedReducedBasis2016}.

\subsection{RB assembly using MPSs}
We now aim to combine the RB framework with tensor network techniques from quantum many-body physics.
More specifically, we represent the snapshot many-body ground states as MPSs 
\begin{equation}
    \ket{\psi} = \!\sum_{\alpha_1 \cdots \alpha_L} \sum_{a_1 \cdots a_{L+1}}
    \!\!\!M^{\alpha_1}_{a_1 a_2} M^{\alpha_2}_{a_2 a_3} \cdots M^{\alpha_L}_{a_L a_{L+1}}
    \ket{\alpha_1 \cdots \alpha_L},
\end{equation}
for systems of $L$ physical degrees of freedom $\alpha_i$.
In doing so, the treatment of large many-body Hamiltonians becomes feasible since efficient algorithms for the computation of state overlaps, matrix elements and ground-state MPSs exist, that scale polynomially in $L$.
These rely on low-rank approximations of the MPS tensors $M^{\alpha_i}_{a_i a_{i+1}}$ which can be implemented by repeated singular value decompositions to reduce the matrix rank $d$, known as the bond dimension in tensor network theory~\cite{orusPracticalIntroductionTensor2014}, by dropping singular values according to a cutoff
\begin{equation} \label{eqn:singular-value-cutoff}
    \cut_\sigma > \sqrt{\frac{\sum_{k \in \mathrm{trunc}} \sigma_k^2}{\sum_{k=1}^{d} \sigma_k^2}}.
\end{equation}
The sum $\sum_{k \in \mathrm{trunc}}$ is to be understood in the sense that we remove the smallest singular values $\sigma_k$ once the singular value error surpasses $\cut_\sigma$, which corresponds to a truncation of the MPS tensors in the Frobenius norm.
In order to perform truth solves in the MPS format, we use DMRG, which is the most commonly used method for variational ground MPS searches.
For reviews on general MPS theory and the DMRG algorithms we refer the reader to Refs.~\cite{Verstraete2008,schollwoeckDensitymatrixRenormalizationGroup2011,orusPracticalIntroductionTensor2014}.

Carrying out the greedy offline stage in MPS formulation spawns two new aspects: First, we need to use efficient MPS contractions for overlaps and matrix elements when computing the reduced quantities $b$, $h$ and $h^2$ which generates additional inaccuracies on top of the truth solve, and second, we need an orthogonalization scheme for $B$ that is compatible with MPS snapshots.
Here, we elaborate on these aspects by going through the greedy algorithm once again, and postpone the discussion of MPS inaccuracies in the RB framework to Sec.~\ref{sec:mps-approx}.

For each iteration of the greedy assembly, we obtain $m_j$ normalized ground-state MPSs and append them to the matrix $\Upsilon = [\ket{\Psi_1} \cdots \ket{\Psi_{M}}] \in \mathbb{C}^{\mathcal{N} \times M}$ as column vectors.
In practice, this matrix cannot be constructed explicitly and only operations between its columns are allowed.
To orthogonalize this matrix, we make the ansatz $B = \Upsilon V$, where $V \in \mathbb{C}^{M \times N}$ mixes the truth MPSs into orthogonal linear combinations. However, these linear combinations are not computed explicitly since this would entail inefficient MPS addition.
Instead, we first compute the overlap matrix
\begin{equation}
    S \coloneqq \Upsilon^\dagger \Upsilon,\quad
    S_{ij} = \braket{\Psi_i | \Psi_j},
\end{equation}
and matrix elements $[\Upsilon^\dagger A \Upsilon]_{ij} = \braket{\Psi_i | A | \Psi_j}$ where $i,j \in \{1,\dots,M\}$ and $A=H_q, H_q H_{q'}, O_r$ might be any operator of interest.
Note that we assume the operator $A$ to be represented in tensor format, e.g., as a matrix product operator (MPO) or a multi-site operator.
Then all reduced quantities are computed in the orthogonal basis by transforming in RB space as
\begin{equation}
    b = V^\dagger S V,\quad
    a = V^\dagger \Upsilon^\dagger A \Upsilon V,
\end{equation}
where correspondingly $a = h_q, h_{qq'}, o_r$.
Note that by adding a new MPS snapshot $\ket{\bm\Psi\ArgI{n+1}}$ to $\Upsilon$, we need to compute $M+m_{n+1}$ new overlaps and matrix elements per observable, exploiting the hermiticity of $S$ and $\Upsilon^\dagger A \Upsilon$ (granted that $A$ is Hermitian).

While there are multiple options to determine the orthogonalizing matrix $V$, we here opt for the numerically efficient approach of L\"owdin symmetric orthogonalization~\cite{Loewdin1950}.
To that end, we decompose the overlap matrix $S = U \Lambda U^{-1}$ into its eigenvalues $\Lambda = \mathrm{diag}(\lambda_1, \dots, \lambda_{M})$, sorted in descending order.
Since $S$ is Hermitian, $U$ can be chosen to be a unitary matrix, so that by demanding $b = V^\dagger S V \overset{!}{=} I$, we can immediately identify
\begin{equation}
    V = U \Lambda^{-1/2}.
\end{equation}
Furthermore, the eigenvalue decomposition of $S$ provides a way to compress the RB.
Due to accumulating linear dependence, the eigenvalues $\lambda_j$ decrease as snapshots are appended until we cannot further resolve new directions in RB space, given the truth solver's accuracy.
Hence we may truncate $B$ in the Frobenius norm of $S$ according to an eigenvalue cutoff
\begin{equation} \label{eqn:lambda-cutoff}
    \cut_\lambda > \sqrt{\frac{\sum_{j \in \mathrm{trunc}} \lambda_j^2}{\sum_{j=1}^{M} \lambda_j^2}},
\end{equation}
i.e., when the smallest normalized sum of squared eigenvalues exceeds $\cut_\lambda$ the corresponding snapshots are removed.
This approach of orthogonalizing a set of snapshots is akin to what is known as \textit{proper orthogonal decomposition}~\cite{quarteroniReducedBasisMethods2016, hesthavenCertifiedReducedBasis2016} in RB theory, which provides an (expensive) alternative to the greedy sampling algorithm.

On a slightly more technical note, we mention the possibility of using the RB to produce initial guesses for the DMRG solver.
By explicitly computing the linear combination $\ket{\Phi^{[i]}_\mathrm{rb}\Arg} = B \varphi^{[i]}_\mathrm{rb}\Arg = \sum_{j=1}^{N} [V \varphi^{[i]}_\mathrm{rb}\Arg]_j \ket{\Psi_j}$ for $i=1,\dots,m$ at the selected parameter point and using it as the initial MPS, one can speed up DMRG convergence.
In order for this to work sufficiently fast, it is necessary to heavily truncate the MPS while computing the linear combination, since MPS addition leads to an additive increase of bond dimensions.
Nonetheless, this approach globally reduces the number of DMRG sweeps and makes the RB assembly more stable as well as deterministic.

Before proceeding with the numerical results, we remark that the procedures we covered here in the context of MPSs generalize to a larger class of vector representations and truth solvers.
In summary, one can identify the following requirements for compatibility with the RB framework:
\begin{enumerate}
    \item Computation of state overlaps $\langle \Psi\ArgI{i} | \Psi\ArgI{j} \rangle$.
    \item Computation of matrix elements $\langle \Psi\ArgI{i} | A | \Psi\ArgI{j} \rangle$ for all relevant operators $A = H_q, H_q H_{q'}, O_r$.
    \item High truth solver accuracy: Large approximation errors on overlaps and matrix elements prevent us from generating a meaningful surrogate. Since this turns out to be a subtle point, we will further discuss this for MPSs in Sec.~\ref{sec:mps-approx}.
\end{enumerate}
In consequence, we do not necessarily need access to the truth ground-state vectors, only to contractions between them.
This opens the door to further tensor network architectures such as projected entangled pair states~\cite{Verstraete2004} and tree tensor networks~\cite{Shi2006,Gerster2014,Gerster2017}, or possibly, artificial neural network based representations~\cite{Caleo2017}
as well as various truth solving methods, e.g., quantum Monte Carlo approaches~\cite{Gubernatis2016}.

\section{Results} \label{sec:results}
We now turn to discuss various numerical results obtained using the RB-MPS method.
Alongside this work, we developed a code package implementing RB methods for parametrized eigenvalue problems, in particular many-body Hamiltonians, with the possibility for using DMRG and ED-based solvers as well as custom truth solving methods.
The code package is written in the Julia programming language~\cite{bezansonJuliaFreshApproach2017} and all MPS and DMRG procedures are performed using the ITensor library~\cite{fishmanITensorSoftwareLibrary2022, fishmanITensor03Release2022}.
We made the code, including user instructions and documentation, publicly available~\cite{GitHubRepo}.

\subsection{Haldane spin-1 chain with single-ion anisotropies} \label{sec:haldane}

\begin{figure}
    \centering
    \includegraphics{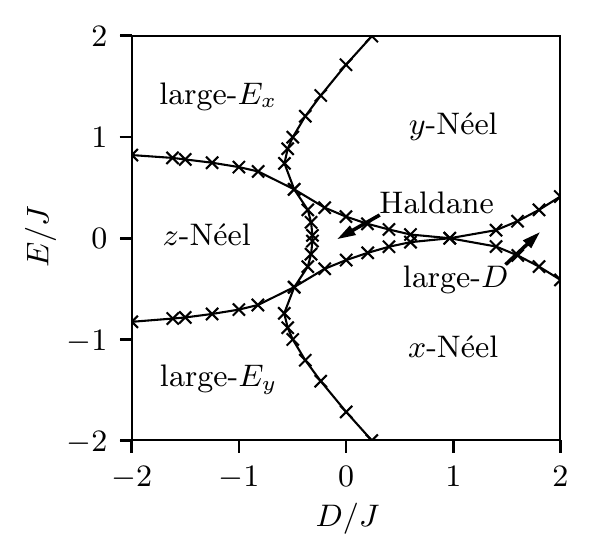}
    \caption{Ground-state phase diagram of the spin-1 chain in Eq.~\eqref{eqn:haldane-chain}, where the phase boundaries are taken from Ref.~\cite{tzengQuantumPhaseTransitions2017}. In the N\'eel phases the model exhibits antiferromagnetic order in $x$, $y$ and $z$-direction, respectively. The Haldane phase is gapped and an example of symmetry-protected topological order and exhibits non-local string order. The uniaxial anisotropic coupling leads to the large-$D$ phase that confines the spins to the $xy$-plane, since the $D$-term penalizes configurations which are polarized in $z$-direction. Similarly, due to the rhombic coupling the large-$E_x$ and $E_y$ phases, which differ in the sign of $E$, favor configurations in the $yz$ and $xz$-plane.}
    \label{fig:haldane-phase-diagram}
\end{figure}

As a first application of the RB-MPS method, we consider the one-dimensional Haldane spin-1 chain
\begin{equation} \label{eqn:haldane-chain}
    H_\mathrm{HD}\! =\! J \sum_{i=1}^{L-1} \bm{S}_i \cdot \bm{S}_{i+1} 
    + D \sum_{i=1}^{L} (S_i^z)^2 + E \sum_{i=1}^{L}\big[ (S_i^x)^2 - (S_i^y)^2 \big],
\end{equation}
with a uniaxial $D$ and a rhombic-type $E$ single-ion anisotropy, here using open boundary conditions.
With regard to the RB formalism we express $H_\mathrm{HD}$ as a dimensionless affine decomposition using the parameter vector $\bm\mu = (D/J,\, E/J)$ and the coefficient function $\bm\theta\Arg = (1, \mu_1, \mu_2)$, such that the matrices $H_q$ correspond to the summands of $H_\mathrm{HD}$.
In order to benchmark the RB-MPS method, we want to resolve the ground-state phase diagram of $H_\mathrm{HD}$, which features: i) the symmetry-protected topological Haldane phase~\cite{Haldane1981,Haldane1983,Haldane1983B,Gu2009,Pollmann2012} that is robust against small anisotropies, ii) N\'eel-ordered phases, as well as so-called iii) large-$D$ and iv) large-$E$ phases.
We summarize and sketch the full phase diagram in Fig.~\ref{fig:haldane-phase-diagram}: cf. Refs.~\cite{tzengQuantumPhaseTransitions2017,Ren2018} for a detailed discussion and an overview of previous investigations. 
DMRG studies of $H_\mathrm{HD}$ can profit from several parity quantum numbers~\cite{Tzeng2012}.
In particular, $H_\mathrm{HD}$ conserves the magnetization parity $p_m = \sum_{i=1}^L S_i^z \!\!\mod 2 \in \{0, 1\}$ together with the spatial parity $p_s=\pm 1$ and time reversal $t=\pm 1$.
The ground-state sector is described by $p_m=0$, whereas the corresponding $p_s$ and $t$ can be determined based on the valence-bond-solid picture~\cite{Affleck1987} in dependence of the boundary conditions~\cite{tzengQuantumPhaseTransitions2017}.
In our RB-MPS approach we thus use DMRG with Abelian quantum number conservation and fix $p_m=0$ to operate in the ground-state sector.
Note that $p_s$ and $t$ are not explicitly fixed here.

For the RB-MPS calculations we consider spin chains of length $L \in \{20, 40, 60, 80\}$ using MPSs with a singular value error cutoff
\footnote{Note that in the ITensor convention, the singular value cutoffs are defined via the squared Frobenius norm, such that the ITensor cutoffs correspond to $\cut_\sigma^2$.}
of up to $\cut_\sigma = 10^{-8}$ --- corresponding to bond dimensions of mostly $d \sim 100$ up to 1000, depending on the phases --- and DMRG sweeps which are converged up to an energy tolerance of $10^{-9}$.
Furthermore, we focus on the upper half-plane $\mathbb{P}_\mathrm{HD} = [-2,2] \times [0,2]$ since the remaining phase diagram is obtained by flipping the sign of $E$ which corresponds to swapping $x$ and $y$ terms.
On this domain, we use a regular training grid of $80 \times 60$ parameter points and converge the surrogate up to an overlap eigenvalue accuracy of at least $\cut_\lambda < 10^{-8}$.
Note that we here and in all following examples do not target degenerate ground-states with DMRG solves, such that the number of snapshots $n = N$ corresponds to the RB dimension.

\begin{figure}
    \centering
    \includegraphics{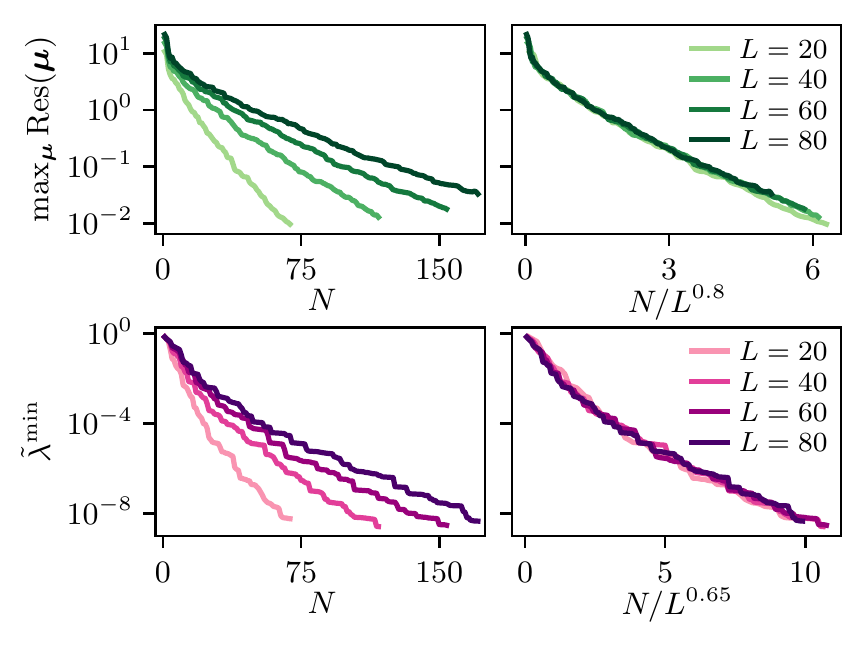}
    \caption{Maximal residual (upper panels) and the normalized minimal eigenvalue $\tilde{\lambda}^\mathrm{min} = \lVert \Lambda \rVert^{-1}_F \min(\Lambda)$ of the overlap matrix $S$ (lower panels) in dependence of the RB dimension $N$. Both quantities follow an exponential decay, although with different rates. The right panels show the curves on rescaled $x$-axes which illustrates the scaling behavior of $N$ with system size. More specifically, we find $N \sim L^\eta$ with approximately $\eta = 0.8$ for the residual and $\eta = 0.65$ for the minimal eigenvalues.}
    \label{fig:haldane-residual-decay}
\end{figure}

\begin{figure}
    \centering
    \includegraphics{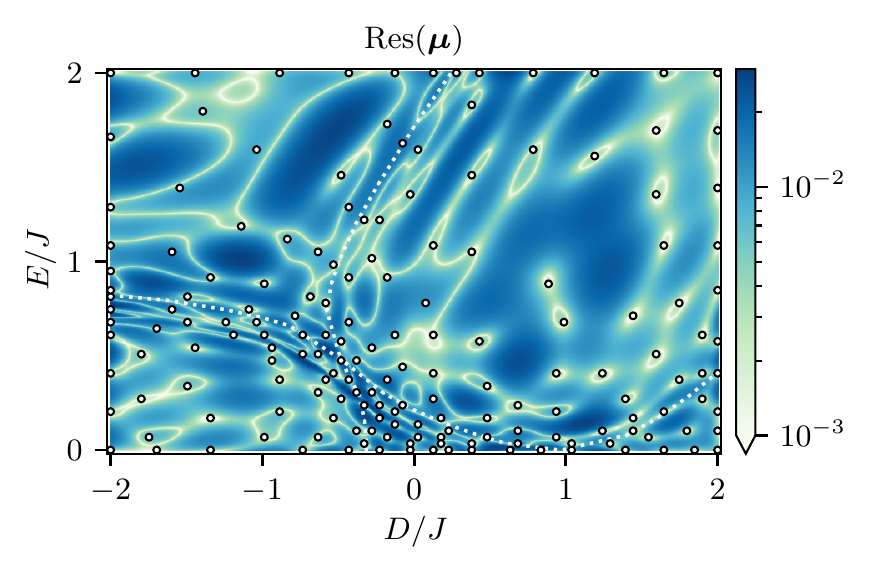}
    \caption{Residual error estimate $\mathrm{Res}\Arg$ on $\mathbb{P}_\mathrm{HD}$. The dots indicate the $N=173$ parameter points $\{\bm\mu_1,\dots,\bm\mu_N\}$ corresponding to the snapshots that span the RB space. The dotted white lines show the phase boundaries of~\cite{tzengQuantumPhaseTransitions2017}.}
    \label{fig:haldane-residual-heatmap}
\end{figure}

At these settings, the greedy algorithm takes $N = 173$ snapshots to reach convergence at the largest system size of $L=80$, with maximal residuals of order $\max_{\bm\mu \in \Xi_\mathrm{train}} \mathrm{Res}\Arg \sim 10^{-2}$.
Only a fraction of the $80 \times 60$ possible training points is thus needed to converge the surrogate model.
To illustrate the convergence, we show the residual as well as the decay of the minimal eigenvalue of the overlap matrix $S$ as a function of the RB dimension $N$ in Fig.~\ref{fig:haldane-residual-decay}.
The exponential decay of the overlap eigenvalues numerically demonstrates the increasing linear dependence among ground states on $\mathbb{P}_\mathrm{HD}$ as snapshots are appended to the RB.
Viewed from a physical angle, the fact that the decrease of overlap eigenvalues coincides with decreasing residuals over the parameter domain can be interpreted as larger eigenvalues being associated with global ground-state behavior, while smaller eigenvalues resolve more localized features in the phase diagram. 
Unsurprisingly, the decay rates decrease with growing system sizes --- in the thermodynamic limit $L \to \infty$ we would in fact run into an orthogonality catastrophe, requiring a diverging number of snapshots to assemble a RB.
Note that this also hinders the immediate use of \textit{infinite} MPSs~\cite{Vidal2007,McCulloch2008,ZaunerStauber2018} in the RB framework, since all overlaps and matrix elements between ground states at different parameter points would vanish, effectively prohibiting any linear combinations between snapshots.
However, for $H_\mathrm{HD}$ on $\mathbb{P}_\mathrm{HD}$, the RB dimension required to converge a surrogate up to a fixed residual increases merely sublinearly in $L$, as opposed to the exponential Hilbert space growth, which indicates that there exists a sweet spot for system sizes, where the thermodynamic regime is approached while still being amenable to the RB-MPS method.

It is also interesting to examine the residual on the entire parameter domain, i.e., on a high-resolution online grid covering $\mathbb{P}_\mathrm{HD}$, together with the snapshot parameter points in Fig.~\ref{fig:haldane-residual-heatmap}.
In particular, many snapshots are selected along phase boundaries and especially around the transition point between the large-$E_x$ and Haldane phase, whereas deep in the phases less snapshots are needed to resolve the ground-state behavior.
We also find that a noticeable number of snapshots are located along the boundary of the parameter range $\mathbb{P}_\mathrm{HD}$, which is standard in greedy RB algorithms.
The clustering of snapshot parameter points hints at the selection mechanism of the greedy algorithm that indicates domains in the phase diagram where the ground states vary more rapidly.
More specifically, a higher density of sample points indicates a slower decay of the local Kolmogorov $N$-width~\cite{Binev2011} and thus a higher linear independence of the solutions under local parameter variation.

\begin{figure}
    \centering
    \includegraphics{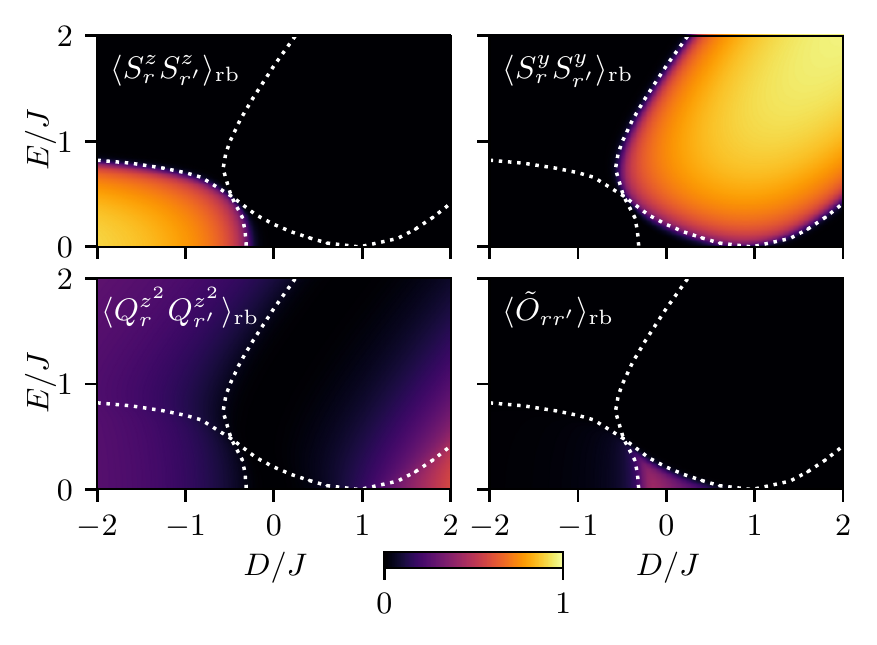}
    \caption{Different two-site correlation functions measured on $\mathbb{P}_\mathrm{HD}$ for an $L=80$ chain at sites $r=20$ and $r'=60$. The dotted white lines show the phase boundaries obtained in~\cite{tzengQuantumPhaseTransitions2017}. In the upper left and right panels, the N\'eel phases are clearly indicated by the spin correlators in $z$ and $y$-direction, whereas the quadrupolar correlator in the bottom left panel is smoothed along the phase boundaries. The bottom right panel shows the modified string order parameter from Eq.~\eqref{eqn:modified-sop} where spin-spin correlations in $z$-direction were subtracted.}
    \label{fig:haldane-correlators}
\end{figure}

In order to reproduce the phase diagram of Fig.~\ref{fig:haldane-phase-diagram}, we next measure various correlation functions that distinguish the different types of ground-state order~\cite{{tzengQuantumPhaseTransitions2017}}.
The N\'eel-ordered phases are identified by measuring the spin-spin correlation functions $\langle S^\alpha_r S^\alpha_{r'} \rangle\rb$ in the $\alpha=z$ and $y$-direction, while the large-$D$ and large-$E_x$ phases are characterized by quadrupolar correlations $\langle Q^\gamma_r Q^\gamma_{r'} \rangle\rb$ where we focus on the $\gamma = z^2$ order with $Q^{z^2}_r = \big[3(S^z_r)^2 - 2\big] / \sqrt{3}$.
To detect the Haldane phase, we consider the non-local string order operator~\cite{DenNijs1989} $O_{rr'} = -S^z_r \prod_{j=r+1}^{r'-1}e^{i \pi S^z_j} S^z_{r'}$ from which we subtract the spin correlator in $z$-direction to remove the trivial background signal in the $z$-N\'eel phase and isolate the string order in the Haldane phase,
\begin{equation} \label{eqn:modified-sop}
    \langle \tilde{O}_{rr'} \rangle\rb =
    -\big\langle S^z_r e^{i \pi \sum_{j=r+1}^{r'-1} S^z_j} S^z_{r'} \big\rangle\rb
    -\langle S^z_r S^z_{r'} \rangle\rb.
\end{equation}
We show the corresponding measurements obtained from the $L=80$ RB in Fig.~\ref{fig:haldane-correlators}.
The RB correctly reproduces the different ground-state phases, up to finite-size effects, matching the results from Ref.~\cite{tzengQuantumPhaseTransitions2017}. 
It can be clearly seen how the string order persists in the Haldane phase also for small values of the anisotropies $D$ and $E$.

Overall, we find that the RB-MPS approach allows us to efficiently uncover the parameter regions of the various ground-state phases at high resolution, based on a comparably low number $N\sim 100$ of DMRG truth solves.
We leave a quantitative treatment of the accuracy of RB measurements to Sec.~\ref{sec:technical-discussion} and move to a more complex example application. 

\subsection{Bilinear-biquadratic spin-1 chain with uniaxial single-ion anisotropy} \label{sec:blbq}

\begin{figure}
    \centering
    \includegraphics{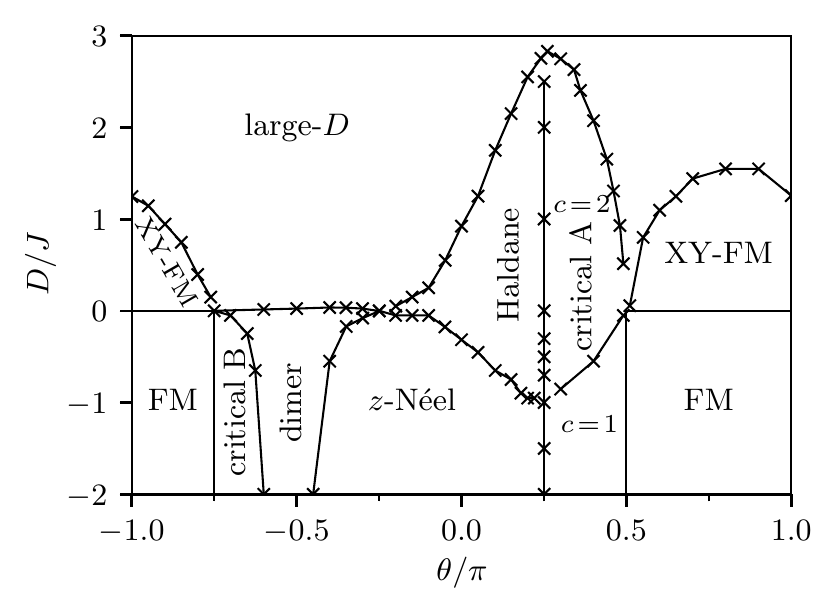}
    \caption{Phase diagram of the bilinear-biquadratic spin-1 chain with uniaxial coupling $D$. The phase boundaries are taken from Ref.~\cite{dechiaraBilinearbiquadraticSpin1Chain2011}, here shown without error bars. The $\theta=0$ cut is equivalent to the $E=0$ cut of $H_\mathrm{HD}$ and therefore already contained in Fig.~\ref{fig:haldane-phase-diagram}. Besides the Haldane, large-$D$ and $z$-N\'eel phase, further regimes are the fully polarized ferromagnetic (FM) and in-plane XY-FM ordered phases, a dimerized phase with a finite bond order parameter (\ref{eq:BOW}), and two critical phases, denoted critical A and critical B. In the former, the central charge $c=1$, while for the latter, two domains have been identified, where $c=1$ and $c=2$, respectively.}
    \label{fig:blbq-phase-diagram}
\end{figure}

In the second example we stay in the realm of spin-1 chains.
We now add a biquadratic exchange interaction term, while removing the rhombic anisotropy,
resulting in the bilinear-biquadratic model with a uniaxial single-ion anisotropy
\begin{align}
    H_\mathrm{BLBQ} =\ &J\sum_{i=1}^{L-1} \big[ \cos(\theta) \bm{S}_i \cdot \bm{S}_{i+1} +
    \sin(\theta) (\bm{S}_i \cdot \bm{S}_{i+1})^2 \big]\nonumber\\
    &+ D \sum_{i=1}^{L} (S^z_i)^2,
\end{align}
on a chain with open boundaries.
Written as a dimensionless affine decomposition, we identify the parameter vector $\bm\mu = (\theta,\, D/J)$ and coefficient function $\bm\theta\Arg = (\cos(\mu_1), \sin(\mu_1), \mu_2)$.
Various aspects of the rather rich physics of this model, featuring multiple gapped as well as critical phases, were previously
studied using both analytical and computational approaches, including extensive DMRG calculations.
We refer in particular to the overall ground-state phase diagram reported in Ref.~\cite{dechiaraBilinearbiquadraticSpin1Chain2011}, which also provides an overview of previous studies.
The ground-state phase diagram of $H_\mathrm{BLBQ}$ is outlined and visualized in Fig.~\ref{fig:blbq-phase-diagram}.
Due to the increased complexity in parameter space, the anisotropic bilinear-biquadratic spin chain serves as an interesting stress test for the RB-MPS approach.

To begin with, we consider the entire parameter domain of Fig.~\ref{fig:blbq-phase-diagram}, corresponding to $\mathbb{P}_\mathrm{BLBQ} = [-\pi, \pi] \times [-2, 3]$, covering all possible phases.
Here, the training grid consists of $100 \times 100$ uniformly spaced parameter points.
Since this gives rise to a significant linear independence between snapshots, we restrict ourselves to a small system of $L=24$ spins.
Again an energy convergence tolerance of $10^{-9}$ is used,
while the singular value cutoff is reduced to $\cut_\sigma = 10^{-5}$, due to increased computational complexity of DMRG in the critical phases.
At this relatively low precision, we face the problem that the induced MPS errors prohibit the greedy algorithm to resolve the phase diagram up to the desired accuracy --- we merely reach a maximal residual of $\max_{\bm\mu \in \Xi_\mathrm{train}} \mathrm{Res}\Arg = 0.2$ at $N=198$ basis snapshots, before terminating the assembly.
Hence the results obtained from the RB at these settings have to be interpreted with caution.
We follow up on the interplay of MPS accuracy and RB convergence in Sec.~\ref{sec:mps-approx}, and here first discuss the numerical results.
Despite the crude accuracy, we are able to make several interesting observations.

\begin{figure}
    \centering
    \includegraphics{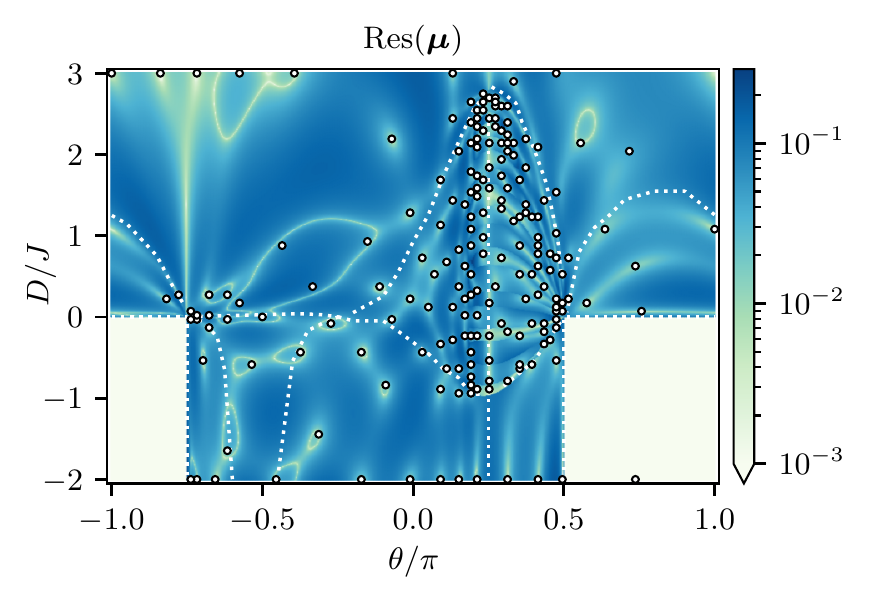}
    \caption{Residual on $\mathbb{P}_\mathrm{BLBQ}$ as well as the $N=198$ selected parameter points for the $L=24$ bilinear-biquadratic model. The critical A phase at central charge $c=2$ accumulates the bulk of the parameter points, whereas the FM domain is emulated by only one snapshot, with the surrogate obtaining a low residual across the FM phase.}
    \label{fig:blbq-residual-heatmap}
\end{figure}

We begin by illustrating the greedy parameter selection, where the snapshot parameters and residual on $\mathbb{P}_\mathrm{BLBQ}$ are shown in Fig.~\ref{fig:blbq-residual-heatmap}.
It can be observed that the bulk of snapshots concentrates around the critical A phase and the $c=2$ region features plateau-like structures in the residual.
Moreover, a less concentrated clustering can be seen around the $\mathrm{SU}(3)$ point~\cite{rodriguezFieldInducedPhaseTransitions2011} at $\bm\mu = (-3/4\pi, 0)$ and the entire ferromagnetic phase is spanned by only one snapshot.

\begin{figure}
    \centering
    \includegraphics{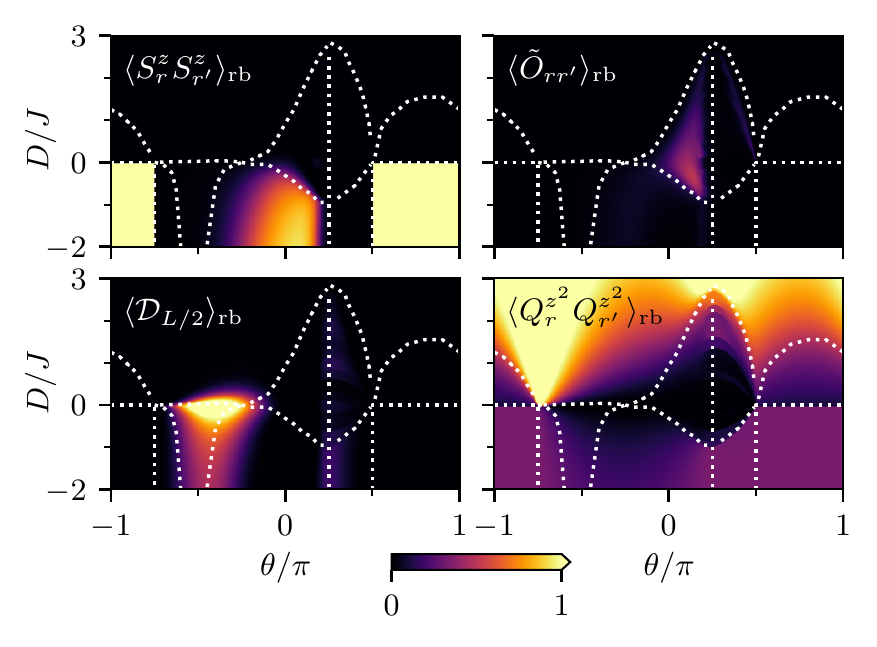}
    \caption{Correlation functions on the entire parameter domain $\mathbb{P}_\mathrm{BLBQ}$. Since we consider only a small $L=24$ spin chain, we observe finite-size effects such as shifted and blurred phase boundaries, e.g., in the dimerized phase. The correlators are measured at sites $r=6$ and $r'=19$, and the dimer order parameter is computed in the chain center. Note that the color bar scale is truncated at 1, which is slightly surpassed by the dimer order parameter and quadrupolar correlator.}
    \label{fig:blbq-correlators}
\end{figure}

We proceed and investigate the gapped domains in parameter space.
As in the previous application, the $z$-N\'eel and large-$D$ phases are again indicated by the $z$-spin and $z^2$-quadrupolar correlation functions, respectively, and the Haldane phase is captured by the appropriate string order parameter $\langle \tilde{O}_{rr'} \rangle\rb$ of Eq.~\eqref{eqn:modified-sop}.
To detect the dimerized phase, we consider the bond order parameter
\begin{equation}\label{eq:BOW}
    \mathcal{D}_r\Arg = |H_r\Arg - H_{r+1}\Arg|,
\end{equation}
where $H_r\Arg = \cos(\theta) \bm{S}_r \cdot \bm{S}_{r+1} + \sin(\theta) (\bm{S}_r \cdot \bm{S}_{r+1})^2$, i.e., the difference of next-neighbor interactions among three adjacent sites.
We present the RB measurements in Fig.~\ref{fig:blbq-correlators}.
The most pronounced signal for dimerization is obtained along the $D=0$ line.
The RB results furthermore indicate that the phase boundary between the dimerized and the large-$D$ regime is located at finite $D>0$ near the $D=0$ line.
This is in accord with earlier high-precision determinations of this phase boundary line~\cite{Fath1995, Buchta2005, Rizzi2005, Laeuchli2006, rodriguezFieldInducedPhaseTransitions2011, dechiaraBilinearbiquadraticSpin1Chain2011, Hu2014, Weyrauch2018}.

Next, we consider the critical phases.
Previous investigations~\cite{rodriguezFieldInducedPhaseTransitions2011, dechiaraBilinearbiquadraticSpin1Chain2011} pointed out that these exhibit dominant (algebraic) correlations of quadrupolar spin-nematic operators.
In order to systematically investigate these correlations, we measured various quadrupolar structure factors
\begin{equation}
    \langle Q^\gamma(-k)\, Q^\gamma(k) \rangle\rb = \frac{1}{L} \sum_{r,r'=1}^{L}
    \!\exp[-ik(r-r')]\, \langle Q^\gamma_r Q^\gamma_{r'} \rangle\rb,
\end{equation}
using the $L=24$ surrogate, with $\gamma \in \{xy, x^2-y^2, z^2\}$, probing thus both transverse and longitudinal quadrupolar correlations, where
$Q^{xy}_r = S^x_r S^y_r + S^y_r S^x_r$, $Q^{x^2-y^2}_r = (S^x_r)^2 - (S^y_r)^2$, and $Q^{z^2}_r$ was given above.
Note that the case $\gamma = x^2-y^2$ is in fact equivalent to the $xy$ case, due to the residual $\mathrm{U}(1)_{xy}$ symmetry of the Hamiltonian $H_\mathrm{BLBQ}$. 
Hence, we do not discuss this case separately here.

\begin{figure}
    \centering
    \includegraphics{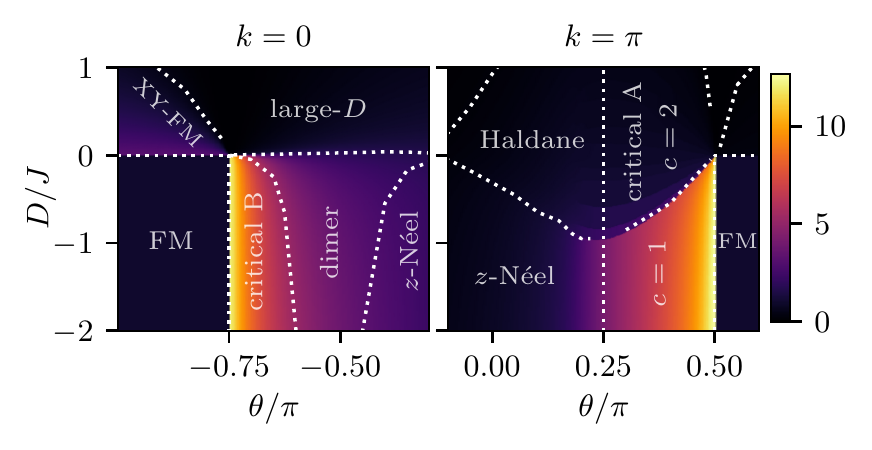}
    \caption{Quadrupolar structure factor $\langle Q^{xy}(-k)\, Q^{xy}(k) \rangle\rb$, shown for $k=0$ in the critical B domain (left) and at $k=\pi$ in the $c=1$ part of the critical A phase (right).}
    \label{fig:blbq-critical-ab}
\end{figure}

The results for $\gamma = xy$ are shown in Fig.~\ref{fig:blbq-critical-ab} in the vicinity of both the critical A and B phases.
From Fig.~\ref{fig:blbq-critical-ab} we detect enhanced transverse ferroquadrupolar correlations, i.e., at $k=0$, in the regime of the critical B phase, as expected~\cite{rodriguezFieldInducedPhaseTransitions2011, dechiaraBilinearbiquadraticSpin1Chain2011}.
Moreover, dominant transverse antiferroquadrupolar correlations with $k=\pi$ are observed in the $c=1$ domain of critical A.
This is in accord with the overall expectation for enhanced nematic correlations in this regime~\cite{dechiaraBilinearbiquadraticSpin1Chain2011}.
The antiferroquadrupolar character has however not been reported in Ref.~\cite{dechiaraBilinearbiquadraticSpin1Chain2011}.
Here, this additional information is directly available from the RB-MPS approach.

\begin{figure}
    \centering
    \includegraphics{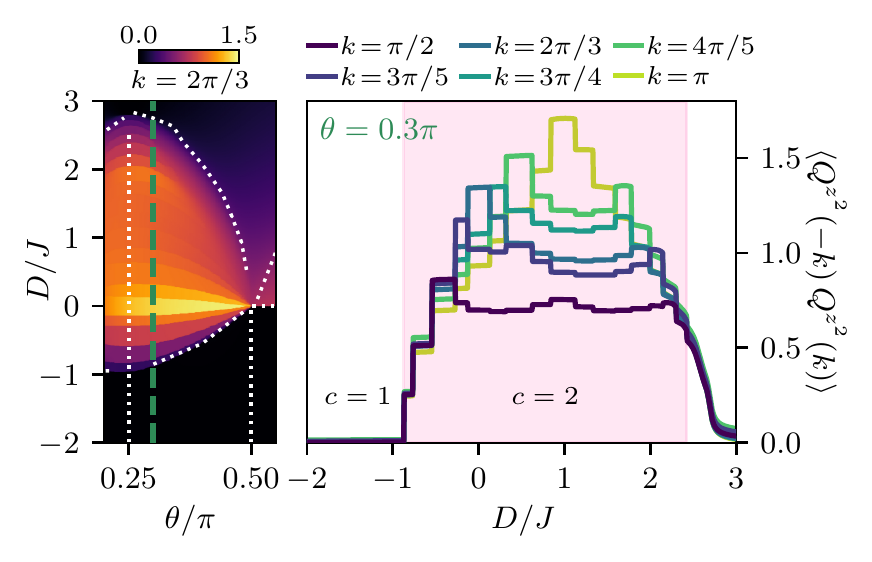}
    \caption{Quadrupolar structure factor $\langle Q^{z^2}(-k)\, Q^{z^2}(k) \rangle\rb$ on the critical A domain at $k=2\pi/3$ (left) and on the parameter cut $\{0.3\pi\} \times [-2, 3]$ (right), which is indicated by the green dashed line. By varying the wave vector, different plateaus are highlighted in the $c=2$ region of the critical phase.}
    \label{fig:blbq-incommensurate}
\end{figure}

For the $\gamma = z^2$ quadrupolar structure factor we observe an enhanced signal in the $c=2$ domain of the critical A phase, as shown in Fig.~\ref{fig:blbq-incommensurate}.
In particular, along the $D=0$ line, the dominant wave vector equals $k=2\pi/3$, as shown in the left panel of Fig.~\ref{fig:blbq-incommensurate}, in accord with earlier reports~\cite{Laeuchli2006}.
For finite values of $D$ however, the signal at $k=2\pi/3$ reduces, and a plateau structure is observed.
The reduced signal could result from either an overall suppression of the quadrupolar correlations or from a shift of the dominant wave vector away from its $D=0$ value of $k=2\pi/3$.
The emergence of the plateau structure indeed already hints towards the latter scenario, as modulations of the spin structure that are not commensurate with the chain length of $L=24$ result in the pinning of the best-matching quadrupolar structure over a finite parameter regime.

In order to further investigate this effect, we narrow down the parameter domain to the one-dimensional parameter cut $\{0.3\pi\} \times [-2, 3]$, which crosses the critical A phase, and on which we generate a second RB for $L=24$ spins.
On this smaller parameter domain we reach a maximal residual of $10^{-2}$ at $N = 42$ snapshots, using a higher MPS accuracy of $\cut_\sigma=10^{-6}$.
Hence, the RB-MPS method this time requires significantly fewer snapshots and simultaneously is able to rectify previous inaccuracies.
Again, we evaluate the $z^2$-quadrupolar structure factor, which is shown for different wave vectors in the right panel of Fig.~\ref{fig:blbq-incommensurate}.
It can be observed that the dominant wave vector $k$ for the quadrupolar correlations varies across the $c=2$ regime of the critical A phase, suggesting a continuously varying $k$ to emerge in the thermodynamic limit. 
We are not aware that these incommensurate quadrupolar correlations have been previously reported for the Hamiltonian $H_\mathrm{BLBQ}$. 
On the other hand, the presence of such incommensurate correlations provides a simple explanation for the particular abundance of snapshots picked by the greedy algorithm throughout the full $c=2$ domain of the critical A phase.
Note that while a relatively large number of snapshots is required to build the RB on this one-dimensional parameter cut, a comparative scan of truth solves would be significantly more expensive since a fine resolution is needed to detect the sharp jumps between plateaus.

From this particular application it becomes apparent, that one of the strengths of the RB-MPS method lies in being able to scan large domains and cheaply vary additional parameters, such as the wave vector, for many different observables with minimal overhead.
This enables us to gain additional insight during a post-processing stage by considering various order parameters on the fly, which would be significantly more involved and computationally expensive by way of truth solving only.
Furthermore, it proves viable to generate surrogates on smaller parameter domains, once a large-scale scan has been performed, in order to improve accuracy and lower the required RB dimension.
Such a procedure of partitioning the parameter space and respectively generating independent RBs could be systematically implemented using the techniques from Refs.~\cite{Eftang2010,Eftang2011,Eftang2012}.
Partitioning approaches could also allow treating larger system sizes for Hamiltonians, such as $H_\mathrm{BLBQ}$, that exhibit complex phase diagrams.

\section{Technical discussion} \label{sec:technical-discussion}
\subsection{Accuracy and convergence} \label{sec:accuracy}
We move on to assess the accuracy of the RB-MPS method.
Similarly to Ref.~\cite{herbstSurrogateModelsQuantum2022}, we quantify the accuracy of RB surrogates by considering the differences between truth and RB measurements and maximizing over the parameter domain under consideration, to obtain the most conservative error estimates.
In order to sample the regions of the parameter domain that have not been selected by the greedy algorithm and truth solved, we introduce a test grid $\Xi_\mathrm{test}$ that has no mutual points with the training grid.
In practice, the test grid is obtained by shifting the training grid by half a grid spacing in each parameter direction.

Taking these considerations into account, we estimate the eigenvalue error by
\begin{equation}
    \mathtt{err}_\mathrm{val} = \max_{\bm{\mu} \in \Xi_\mathrm{test}} \frac{|E\Arg - E\rb\Arg|}{|E\Arg|},
\end{equation}
i.e., the maximal relative difference between truth $E\Arg$ and RB energies $E\rb\Arg$.
For errors of observables, we consider absolute differences
\begin{equation}
    \mathtt{err}_\mathrm{obs} = \max_{\bm{\mu} \in \Xi_\mathrm{test}} |\langle O\Arg \rangle - \langle O\Arg \rangle\rb|,
\end{equation}
since for the observables that we will treat in the following --- spin-spin correlators and quadrupolar structure factors --- the measured values become exactly zero on certain subdomains.
Note that the absolute errors in these cases provide comparable error quantifiers alongside the relative errors because the range of observable values concentrates around $\langle O\Arg \rangle\rb \sim 0.1$ and $\sim 1$.
The vector error estimate is more subtle since it has to account for different global phases between truth and RB solutions.
We avoid such phase issues by computing the difference of outer products
\begin{equation} \label{eqn:vec-error}
    \delta_\mathrm{vec}\Arg = 
    \frac{\lVert \bm{\Psi}\Arg \bm{\Psi}^\dagger\Arg - \bm{\Phi}\rb\Arg \bm{\Phi}^\dagger\rb\Arg \rVert_F}{\lVert \bm{\Psi}\Arg \bm{\Psi}^\dagger\Arg \rVert_F},
\end{equation}
where we denote $\bm\Psi\Arg \equiv (\ket{\Psi^{[1]}\Arg}, \dots, \ket{\Psi^{[m]}\Arg})$ to declutter notation, and again maximize over the test grid, $\mathtt{err}_\mathrm{vec} = \max_{\bm\mu \in \Xi_\mathrm{test}} \delta_\mathrm{vec} \Arg$.
In this particular form, we would need to explicitly reconstruct Hilbert space dimensional vectors from MPSs, which is exponentially hard in the system size.
This can be circumvented by computing the norm using the Frobenius inner product and the cyclic property of the trace, producing the expression
\begin{equation}
    \delta_\mathrm{vec}\Arg = \sqrt{
    1 + \frac{\lVert \bm\Phi\rb^\dagger\Arg \bm\Phi\rb\Arg \rVert^2_F
    - 2 \lVert \bm{\Phi}\rb^\dagger\Arg \bm{\Psi}\Arg \rVert^2_F}
    {\lVert \bm{\Psi}^\dagger\Arg \bm{\Psi}\Arg \rVert^2_F}},
\end{equation}
where $\lVert \bm\Phi^\dagger\rb\Arg \bm\Phi\rb\Arg \rVert_F = \lVert \bm{\varphi}^\dagger\rb\Arg b \bm{\varphi}\rb\Arg \rVert_F$ and the norm of RB and truth overlaps is found to be
\begin{align}
    \lVert \bm{\Phi}\rb^\dagger\Arg \bm{\Psi}\Arg \rVert^2_F
    &= \sum_{i,j=1}^{m} \big|\! \braket{\Phi_\mathrm{rb}^{[i]}\Arg | \Psi^{[j]}\Arg}\! \big|^2\nonumber\\
    &= \sum_{i,j=1}^{m} \Big| \sum_{k=1}^N [V\varphi_\mathrm{rb}^{[i]}\Arg]_k \braket{\Psi_k | \Psi^{[j]}\Arg}\! \Big|^2.
\end{align}
Thus, we are able to compute the eigenvector error using only RB coefficients and MPS overlaps at the particular parameter point $\bm\mu$, without resorting to $\mathcal{N}$-dependent operations.

\begin{figure}
    \centering
    \includegraphics{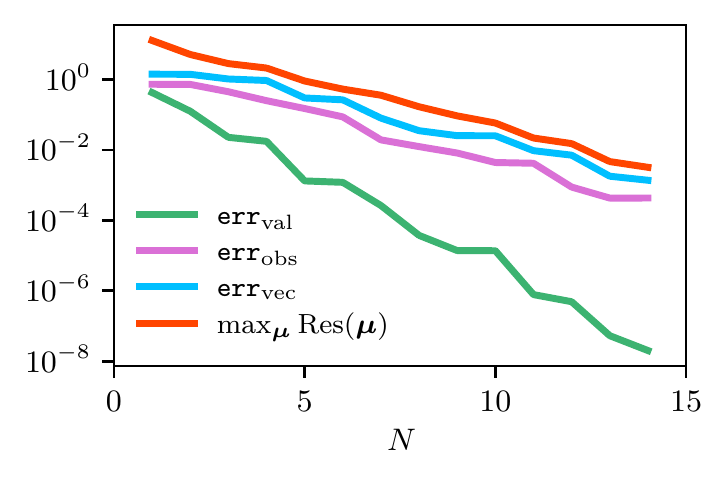}
    \caption{Decay of maximal RB errors for the $L=40$ Haldane chain with respect to the $\{0\} \times [0, 2]$ parameter cut.}
    \label{fig:haldane-error-convergence}
\end{figure}

For the first example, we return to the Haldane chain of $L=40$ spins and compute the above error quantities on the one-dimensional cut of parameters $\bm\mu \in \{0\} \times [0, 2]$.
Note that we restrict ourselves to a medium chain length and small parameter cut due to the high computational effort associated with the required truth solves on $\Xi_\mathrm{test}$.
At a singular value cutoff of $\cut_\sigma = 10^{-8}$ and $N=14$ snapshots, we converge the surrogate to a maximal residual below $10^{-2}$.
For the observable error, we measure the spin-spin correlator $\langle S^y_r S^y_{r'} \rangle$.
The resulting convergence of maximal errors as the RB is generated is shown in Fig.~\ref{fig:haldane-error-convergence}.
It is observed, that the residual decays at a similar rate to the eigenvector error, while the eigenvalue errors show a significantly faster convergence, as was previously found for ED-based surrogate models~\cite{herbstSurrogateModelsQuantum2022}.
Moreover, the data demonstrates that the residual acts as an error surrogate by providing an upper bound to all real errors --- at a residual of $10^{-2}$, the considered error quantities have decayed to the sub-percent range.

\begin{figure}
    \centering
    \includegraphics{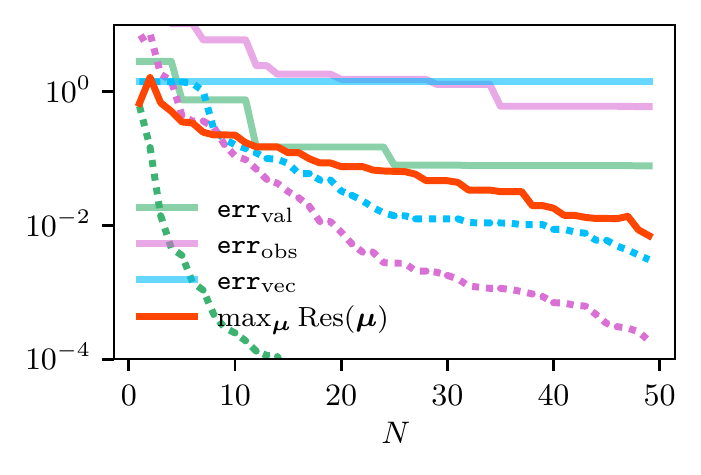}
    \caption{RB errors for the $L=24$ bilinear-biquadratic model on $\bm\mu \in \{0.3\pi\} \times [-2, 3]$. The solid lines show the maximal errors while dotted lines indicate the median errors with respect to $\Xi_\mathrm{test}$.}
    \label{fig:blbq-error-convergence}
\end{figure}

\begin{figure}
    \centering
    \includegraphics{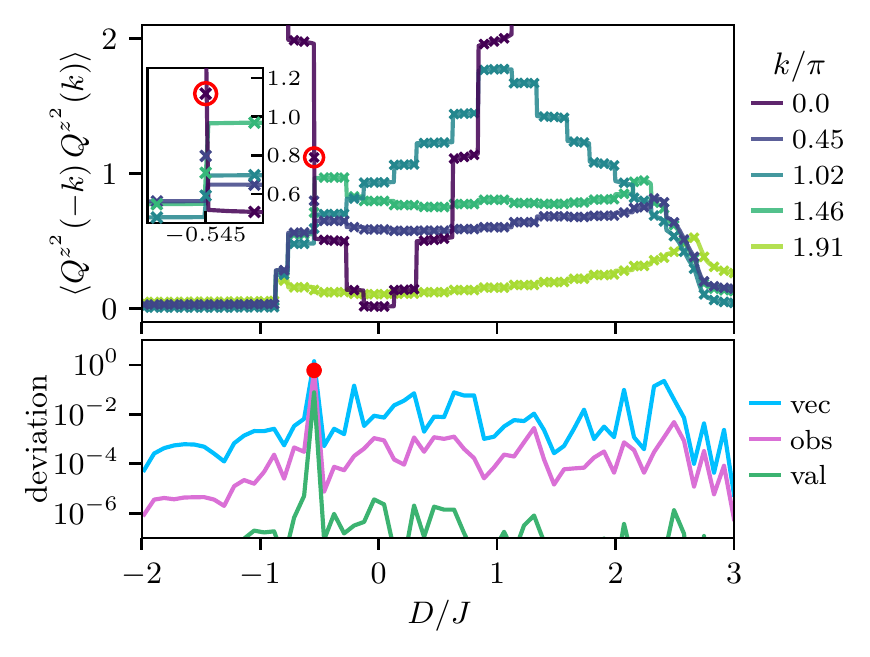}
    \caption{Structure factor $\langle Q^{z^2}(-k)\, Q^{z^2}(k) \rangle\rb$ on $\bm\mu \in \{0.3\pi\} \times [-2, 3]$ computed using the surrogate model (solid lines) and DMRG truth solves on $\Xi_\mathrm{test}$ (markers). The corresponding deviations between DMRG and RB ground states (vec), structure factor measurements (obs) and energies (val) are shown below. Note that we take the maximal structure factor deviations with respect to $k \in [0, 2\pi]$. The maximal observable error, indicated by the red dot, occurs at a particularly error-sensitive transition between two plateaus in the $c=2$ domain, as shown in the inset, where the maximal deviation is produced by the jump in the $k=0$ structure factor (encircled in red).}
    \label{fig:blbq-cut-compare}
\end{figure}

For the next example, we reexamine the $L=24$ bilinear-biquadratic chain on the $\{0.3\pi\} \times [-2, 3]$ parameter cut crossing the critical A phase.
On this parameter domain the surrogate has to resolve a large degree of ground-state variation, and in particular discrete plateaus in the evaluated observables, which serves as a good example to illustrate possible difficulties in converging RBs.
We adopt the offline settings from Sec.~\ref{sec:blbq} and again perform truth solves on a shifted test grid.
As for the observable error, we here consider the quadrupolar structure factor $\langle Q^{z^2}(-k)\, Q^{z^2}(k) \rangle\rb$ from before.
In addition to maximizing the error over the test grid, we also maximize with respect to all possible wave vectors $k \in [0, 2\pi]$.
The maximal and median errors as functions of the RB dimension are presented in Fig.~\ref{fig:blbq-error-convergence}.
Most strikingly, we observe that the maximal RB errors plateau above the maximal residual, which still decays to $10^{-2}$.
While this breaks the desired property of the residual error estimator of providing an upper bound on all RB errors, we see that the median errors do converge nicely below the residual line.
This disparity between maximal and median errors is owed to the discrete jumps observed in the $c=2$ critical A phase as well as the sensitivity of the Fourier transform in the structure factor to variations of $k$.
As shown in Fig.~\ref{fig:blbq-cut-compare}, at one particular point in $\Xi_\mathrm{test}$ a truth solve is performed directly at the transition between two plateaus, where the RB linearly interpolates between the plateau values instead of producing a sharp gap, such that we obtain structure factor deviations of order $10^{-1}$.
Leaving this point aside, the DMRG and RB observable measurements are in excellent agreement, whereas the decay rate of the vector errors is generally slower.

At the chosen MPS accuracies, we thus observe mostly well-behaved convergence properties similar to those of ED-generated RBs, and the fact that the surrogate operates based on approximate MPS snapshots and contractions thereof does not hinder us from obtaining accurate observable measurements.
This prompts the question of when the MPS approximations do become significant and thereby impede the generation of a meaningful surrogate, for which we turn to the next section.

\subsection{Sensitivity to MPS approximations} \label{sec:mps-approx}
One property of RBs which is independent of the specific truth solving method, is that the surrogate accuracy is ultimately bounded by the truth accuracy.
More specifically, when assembling a RB using the greedy algorithm, we expect the residual error estimator to stagnate at some value, since finite truth solving accuracy implies that further snapshots cannot improve the RB accuracy.
In this regard, using DMRG in conjunction with RB methods does not bring forth anything new since ED solvers are also numerically approximate, albeit to higher accuracy.
What is new, however, is that contractions of snapshots with operators $\langle \Psi_i | A | \Psi_j \rangle$ are approximate because applications of MPOs multiplicatively increase bond dimensions, therefore requiring further MPS compressions.

\begin{figure}
    \centering
    \includegraphics{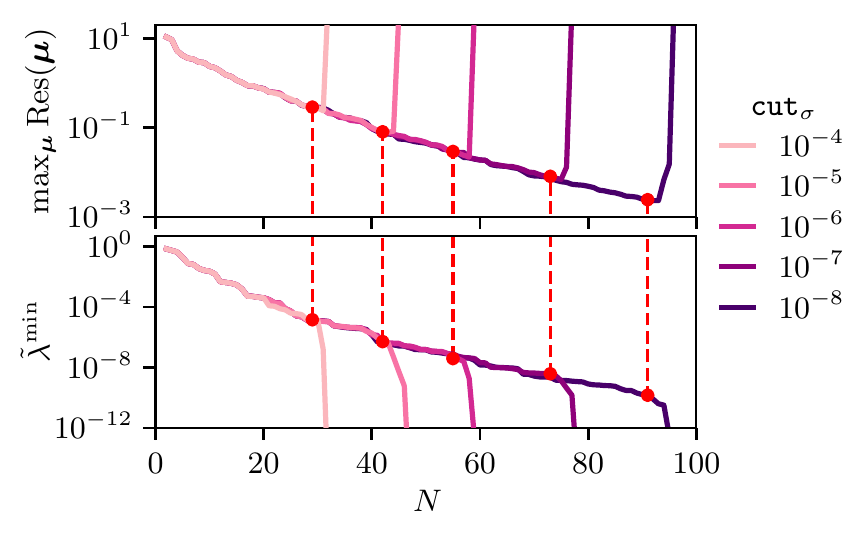}
    \caption{Decay of maximal residuals and minimal normalized overlap eigenvalues $\tilde{\lambda}^\mathrm{min} = \lVert \Lambda \rVert^{-1}_F \min(\Lambda)$ for different singular value error cutoffs $\cut_\sigma$. The red dots indicate the final chosen dimension of the RB surrogates.}
    \label{fig:cutoff-dependency}
\end{figure}

To numerically probe the effects of MPS approximations on the greedy RB assembly, we generate surrogates of the $L=20$ Haldane chain at different singular value cutoffs on the $\mathbb{P}_\mathrm{HD}$ domain, while fixing all other settings.
In particular, we set the residual tolerance and $\cut_\lambda$ to zero, such that the basis assembly could continue indefinitely, and only terminate when reaching a predefined maximal number of snapshots. 
We show the corresponding decay of maximal residuals and minimal overlap eigenvalues in Fig.~\ref{fig:cutoff-dependency}.
Generally it is observed that lower bond dimension cutoffs allow us to assemble surrogates with lower final residuals, so to higher overall accuracy.
Furthermore, for each greedy assembly we encounter a point where the residuals suddenly spike and the minimal eigenvalues drop to zero --- which would be unexpected for ED-based surrogates, where instead the residual is expected to plateau when reaching maximal surrogate accuracy.
The overlap eigenvalues vanish since the greedy algorithm starts to select close-by or even equal parameter points over the span of the last iterations, causing a strong increase in linear dependence.
This is accompanied by approximation errors in $h_q$, $h_{qq'}$ and hence $\bm\varphi\rb$ that eventually prohibit the use of the Rayleigh-Ritz method as in Eqs.~(\ref{eqn:rb-evp}, \ref{eqn:residual-cheap}, \ref{eqn:rb-expectation}) and result in abrupt increases of the residual.
Note that, due to erroneous coefficients $\bm\varphi\rb$ that occur when, and sometimes right before, spiking residuals are produced, RB expectation values of observables then may exhibit unphysical artifacts.

Fortunately, in numerical practice, one can devise simple heuristics to terminate the greedy assembly at the appropriate time, i.e., right before approaching a residual spike (see red dots in Fig.~\ref{fig:cutoff-dependency}), in case sufficient MPS accuracy cannot be guaranteed.
Whenever a parameter point is selected twice, which is forbidden in a correctly assembled greedy basis, the iteration is stopped.
Moreover, a suddenly ill-conditioned RB normalization $b = B^\dagger B$ indicates the drop of minimal overlap eigenvalues, such that the assembly is stopped when $\lVert I - b \rVert > \delta$ for some threshold $\delta$ or equivalently when the condition number of $b$ becomes too large to converge the eigenvalue problem of Eq.~\eqref{eqn:rb-evp}.
We want to remark that even when too many snapshots are included in the RB, it is possible, without further computational expense, to roll back the surrogate model by removing the last snapshots and the corresponding overlaps and matrix elements.
For a more practical discussion of these technical aspects, exemplified by further physical models, we point the reader to our code documentation available via the aforementioned repository~\cite{GitHubRepo}. 

\section{Conclusions} \label{sec:conclusions}
In the present work, we expanded the RB method for quantum spin systems with greedy parameter selection, introduced in Ref.~\cite{herbstSurrogateModelsQuantum2022}, to operate on MPSs obtained from DMRG solves.
MPS approximations allow us to access larger many-body systems than with ED, and thereby open up the RB approach to more physically interesting scenarios, while simultaneously providing sufficient truth accuracy to generate accurate surrogate models.
In order to combine the RB framework with MPS snapshots, we adjusted the orthogonalization method for the reduced basis $B$, and obtained the required overlaps and matrix elements for the RB construction from efficient MPS procedures --- beyond that the RB-MPS approach treats the DMRG as a black-box solver (with finite accuracy), thus profiting from the complementary structure of RB methods.
The combined RB-MPS method was then applied to one-dimensional spin models with rich phase diagrams: the Haldane spin-1 chain with uniaxial and rhombic-type single-ion anisotropies as well as the bilinear-biquadratic spin-1 chain with uniaxial anisotropy.
In both applications, the RB-MPS method proved to be a numerically efficient and versatile tool for determining complex ground-state phase diagrams.

In particular, the phase diagram of the anisotropic Haldane spin-1 chain was efficiently obtained from RB surrogates, where the number of required snapshots grows sublinearly in the system size $L$.
Here, the degree of linear independence across the parameter domain was monitored and controlled by the eigenvalues of the overlap matrix.
The greedy parameter selection mechanism was found to pick up on physical features of the phase diagram: e.g., the Haldane phase of $H_\mathrm{HD}$ was sampled more frequently, whereas the ferromagnetic phase in the bilinear-biquadratic chain model $H_\mathrm{BLBQ}$ was spanned by merely one snapshot.
Therefore, we actually gain an automatized indicator for regions of strong ground-state variation.
Furthermore, it was demonstrated that the RB-MPS approach functions on comparatively large parameter domains containing multiple phases.
Overall, RBs of dimension $N \sim 100$ were sufficient to capture the whole variety of correlations with fairly high resolution.

Especially in the application to $H_\mathrm{BLBQ}$, the RB-MPS approach qualitatively captured a rather rich phase diagram from large-scale parameter scans.
Furthermore, based on a systematic exploration of quadrupolar structure factors within the RB-MPS approach, we were able to identify dominant antiferroquadrupolar correlations within the $c=1$ regime of the critical A phase, and obtained characteristic signatures of incommensurate spin-nematic correlations within the $c=2$ domain of the critical A phase.
Both these findings have apparently not been uncovered by previous investigations, emphasizing the usefulness of a computational tool that efficiently scans large parameter domains.

In the last part, we analyzed the convergence properties of the RB-MPS approach.
For this purpose, we quantified the maximal RB errors and found well-behaved error convergence properties for $H_\mathrm{HD}$, i.e., exponentially decaying RB errors that enter the sub-percent range after $N \sim 10$ snapshots, where the residual constitutes a surrogate error estimator by providing an upper bound on all RB errors.
For the bilinear-biquadratic chain on the critical A phase, the errors converge mostly well across the parameter domain despite the rapid and discrete ground-state changes, except for sharp transitions between structure factor plateaus that are sometimes inaccurately interpolated by the surrogate models.
Lastly, it was explored how the RB accuracy ultimately depends on the chosen MPS singular value cutoff, where we found that the greedy algorithm reaches a point of breaking down that can, however, be avoided by simple heuristics and reversed without additional computational effort.

The presence of MPS approximations does raise issues that depend on the specific physics underlying the system on the sampled domain.
On very elaborate parameter domains with highly linearly independent ground states, i.e., a slowly decaying Kolmogorov $N$-width, the number of required MPS snapshots may become unwieldy --- storing the entire RB in memory hence makes the greedy assembly particularly memory-intensive.
Additionally, we may encounter, e.g., critical phases that necessitate higher bond dimensions and thereby make the DMRG solves considerably more expensive.
Yet, these issues can be circumvented by first performing large-scale scans using a crude RB surrogate and then, informed by the qualitative results, assembling RBs on smaller parameter subdomains with higher accuracy. 
In this sense, the RB-MPS approach offers a particularly convenient workflow for the exploration of unknown phase diagrams: based on a given surrogate, refined RB models can be devised in any relevant subregion of the parameter space upon adapting the offline sampling domain.
Moreover, such subregions can be identified based on readily accessible observables during the online stage at low computational costs.
As one scales up the RB-MPS method to larger system sizes and possibly higher dimensional parameter spaces, such strategies of incremental exploration may become increasingly relevant.

Potential future directions of research on the RB-MPS approach include (automated) grid refinements~\cite{Eftang2010,Eftang2011,Eftang2012} and parallelization of the offline stage (parallelizing the online stage is straightforward).
Furthermore, we anticipate generic ground-state probes such as wave-function overlaps~\cite{Zanardi2006} and associated fidelity susceptibilities~\cite{Gu2007,Gu2010}, as well as entanglement measures~\cite{Eisert2010,Zeng2019} and fluctuations of conserved quantities~\cite{Song2010,Song2012,Rachel2012,Contessi2023}, to be promising further diagnostic tools to combine with the RB-MPS approach presented here.
Finally, we could foresee other tensor network techniques~\cite{Verstraete2004,Shi2006,Gerster2014,Gerster2017} (at least those for finite-size systems) to be naturally integrated in the RB framework, thus extending the application realm to higher spatial dimensions, too.

\begin{acknowledgments}
    We thank Norbert Schuch for helpful suggestions regarding the computation of eigenvector errors using MPSs.
    Simulations were performed with computing resources granted by RWTH Aachen University under project thes1253.
    M.R. acknowledges support from the Deutsche Forschungsgemeinschaft (DFG), project grant 277101999, within the CRC network TR 183 (sub- project B01).
    P.B. and S.W. acknowledge support by DFG through  RTG 1995.
\end{acknowledgments}

% BIBLIOGRAPHY
\bibliography{references}

%apsrev4-2.bst 2019-01-14 (MD) hand-edited version of apsrev4-1.bst
%Control: key (0)
%Control: author (8) initials jnrlst
%Control: editor formatted (1) identically to author
%Control: production of article title (0) allowed
%Control: page (0) single
%Control: year (1) truncated
%Control: production of eprint (0) enabled
\begin{thebibliography}{78}%
\makeatletter
\providecommand \@ifxundefined [1]{%
 \@ifx{#1\undefined}
}%
\providecommand \@ifnum [1]{%
 \ifnum #1\expandafter \@firstoftwo
 \else \expandafter \@secondoftwo
 \fi
}%
\providecommand \@ifx [1]{%
 \ifx #1\expandafter \@firstoftwo
 \else \expandafter \@secondoftwo
 \fi
}%
\providecommand \natexlab [1]{#1}%
\providecommand \enquote  [1]{``#1''}%
\providecommand \bibnamefont  [1]{#1}%
\providecommand \bibfnamefont [1]{#1}%
\providecommand \citenamefont [1]{#1}%
\providecommand \href@noop [0]{\@secondoftwo}%
\providecommand \href [0]{\begingroup \@sanitize@url \@href}%
\providecommand \@href[1]{\@@startlink{#1}\@@href}%
\providecommand \@@href[1]{\endgroup#1\@@endlink}%
\providecommand \@sanitize@url [0]{\catcode `\\12\catcode `\$12\catcode
  `\&12\catcode `\#12\catcode `\^12\catcode `\_12\catcode `\%12\relax}%
\providecommand \@@startlink[1]{}%
\providecommand \@@endlink[0]{}%
\providecommand \url  [0]{\begingroup\@sanitize@url \@url }%
\providecommand \@url [1]{\endgroup\@href {#1}{\urlprefix }}%
\providecommand \urlprefix  [0]{URL }%
\providecommand \Eprint [0]{\href }%
\providecommand \doibase [0]{https://doi.org/}%
\providecommand \selectlanguage [0]{\@gobble}%
\providecommand \bibinfo  [0]{\@secondoftwo}%
\providecommand \bibfield  [0]{\@secondoftwo}%
\providecommand \translation [1]{[#1]}%
\providecommand \BibitemOpen [0]{}%
\providecommand \bibitemStop [0]{}%
\providecommand \bibitemNoStop [0]{.\EOS\space}%
\providecommand \EOS [0]{\spacefactor3000\relax}%
\providecommand \BibitemShut  [1]{\csname bibitem#1\endcsname}%
\let\auto@bib@innerbib\@empty
%</preamble>
\bibitem [{\citenamefont {Haldane}(1981)}]{Haldane1981}%
  \BibitemOpen
  \bibfield  {author} {\bibinfo {author} {\bibfnamefont {F.~D.~M.}\
  \bibnamefont {Haldane}},\ }\href@noop {} {\bibinfo {title} {{Ground State
  Properties of Antiferromagnetic Chains with Unrestricted Spin: Integer Spin
  Chains as Realisations of the O(3) Non-Linear Sigma Model}}} (\bibinfo {year}
  {1981}),\ \bibinfo {note} {{ILL} preprint SP-81/95},\ \Eprint
  {https://arxiv.org/abs/1612.00076} {arXiv:1612.00076} \BibitemShut {NoStop}%
\bibitem [{\citenamefont {Haldane}(1983{\natexlab{a}})}]{Haldane1983}%
  \BibitemOpen
  \bibfield  {author} {\bibinfo {author} {\bibfnamefont {F.~D.~M.}\
  \bibnamefont {Haldane}},\ }\bibfield  {title} {\bibinfo {title} {{Nonlinear
  Field Theory of Large-Spin Heisenberg Antiferromagnets: Semiclassically
  Quantized Solitons of the One-Dimensional Easy-Axis N\'eel State}},\ }\href
  {https://doi.org/10.1103/PhysRevLett.50.1153} {\bibfield  {journal} {\bibinfo
   {journal} {Phys. Rev. Lett.}\ }\textbf {\bibinfo {volume} {50}},\ \bibinfo
  {pages} {1153} (\bibinfo {year} {1983}{\natexlab{a}})}\BibitemShut {NoStop}%
\bibitem [{\citenamefont {Haldane}(1983{\natexlab{b}})}]{Haldane1983B}%
  \BibitemOpen
  \bibfield  {author} {\bibinfo {author} {\bibfnamefont {F.~D.~M.}\
  \bibnamefont {Haldane}},\ }\bibfield  {title} {\bibinfo {title} {Continuum
  dynamics of the {1-D Heisenberg} antiferromagnet: Identification with the
  {O(3)} nonlinear sigma model},\ }\href
  {https://doi.org/https://doi.org/10.1016/0375-9601(83)90631-X} {\bibfield
  {journal} {\bibinfo  {journal} {Phys. Lett. A}\ }\textbf {\bibinfo {volume}
  {93}},\ \bibinfo {pages} {464} (\bibinfo {year}
  {1983}{\natexlab{b}})}\BibitemShut {NoStop}%
\bibitem [{\citenamefont {Gu}\ and\ \citenamefont {Wen}(2009)}]{Gu2009}%
  \BibitemOpen
  \bibfield  {author} {\bibinfo {author} {\bibfnamefont {Z.-C.}\ \bibnamefont
  {Gu}}\ and\ \bibinfo {author} {\bibfnamefont {X.-G.}\ \bibnamefont {Wen}},\
  }\bibfield  {title} {\bibinfo {title} {Tensor-entanglement-filtering
  renormalization approach and symmetry-protected topological order},\ }\href
  {https://doi.org/10.1103/PhysRevB.80.155131} {\bibfield  {journal} {\bibinfo
  {journal} {Phys. Rev. B}\ }\textbf {\bibinfo {volume} {80}},\ \bibinfo
  {pages} {155131} (\bibinfo {year} {2009})}\BibitemShut {NoStop}%
\bibitem [{\citenamefont {Pollmann}\ \emph {et~al.}(2012)\citenamefont
  {Pollmann}, \citenamefont {Berg}, \citenamefont {Turner},\ and\ \citenamefont
  {Oshikawa}}]{Pollmann2012}%
  \BibitemOpen
  \bibfield  {author} {\bibinfo {author} {\bibfnamefont {F.}~\bibnamefont
  {Pollmann}}, \bibinfo {author} {\bibfnamefont {E.}~\bibnamefont {Berg}},
  \bibinfo {author} {\bibfnamefont {A.~M.}\ \bibnamefont {Turner}},\ and\
  \bibinfo {author} {\bibfnamefont {M.}~\bibnamefont {Oshikawa}},\ }\bibfield
  {title} {\bibinfo {title} {Symmetry protection of topological phases in
  one-dimensional quantum spin systems},\ }\href
  {https://doi.org/10.1103/PhysRevB.85.075125} {\bibfield  {journal} {\bibinfo
  {journal} {Phys. Rev. B}\ }\textbf {\bibinfo {volume} {85}},\ \bibinfo
  {pages} {075125} (\bibinfo {year} {2012})}\BibitemShut {NoStop}%
\bibitem [{\citenamefont {Tzeng}\ \emph {et~al.}(2017)\citenamefont {Tzeng},
  \citenamefont {Onishi}, \citenamefont {Okubo},\ and\ \citenamefont
  {Kao}}]{tzengQuantumPhaseTransitions2017}%
  \BibitemOpen
  \bibfield  {author} {\bibinfo {author} {\bibfnamefont {Y.-C.}\ \bibnamefont
  {Tzeng}}, \bibinfo {author} {\bibfnamefont {H.}~\bibnamefont {Onishi}},
  \bibinfo {author} {\bibfnamefont {T.}~\bibnamefont {Okubo}},\ and\ \bibinfo
  {author} {\bibfnamefont {Y.-J.}\ \bibnamefont {Kao}},\ }\bibfield  {title}
  {\bibinfo {title} {Quantum phase transitions driven by rhombic-type
  single-ion anisotropy in the {$S=1$ Haldane} chain},\ }\href
  {https://doi.org/10.1103/PhysRevB.96.060404} {\bibfield  {journal} {\bibinfo
  {journal} {Phys. Rev. B}\ }\textbf {\bibinfo {volume} {96}},\ \bibinfo
  {pages} {060404} (\bibinfo {year} {2017})}\BibitemShut {NoStop}%
\bibitem [{\citenamefont {Ren}\ \emph {et~al.}(2018)\citenamefont {Ren},
  \citenamefont {Wang},\ and\ \citenamefont {You}}]{Ren2018}%
  \BibitemOpen
  \bibfield  {author} {\bibinfo {author} {\bibfnamefont {J.}~\bibnamefont
  {Ren}}, \bibinfo {author} {\bibfnamefont {Y.}~\bibnamefont {Wang}},\ and\
  \bibinfo {author} {\bibfnamefont {W.-L.}\ \bibnamefont {You}},\ }\bibfield
  {title} {\bibinfo {title} {Quantum phase transitions in spin-1 {XXZ} chains
  with rhombic single-ion anisotropy},\ }\href
  {https://doi.org/10.1103/PhysRevA.97.042318} {\bibfield  {journal} {\bibinfo
  {journal} {Phys. Rev. A}\ }\textbf {\bibinfo {volume} {97}},\ \bibinfo
  {pages} {042318} (\bibinfo {year} {2018})}\BibitemShut {NoStop}%
\bibitem [{\citenamefont {Hesthaven}\ \emph {et~al.}(2016)\citenamefont
  {Hesthaven}, \citenamefont {Rozza},\ and\ \citenamefont
  {Stamm}}]{hesthavenCertifiedReducedBasis2016}%
  \BibitemOpen
  \bibfield  {author} {\bibinfo {author} {\bibfnamefont {J.~S.}\ \bibnamefont
  {Hesthaven}}, \bibinfo {author} {\bibfnamefont {G.}~\bibnamefont {Rozza}},\
  and\ \bibinfo {author} {\bibfnamefont {B.}~\bibnamefont {Stamm}},\ }\href
  {https://doi.org/10.1007/978-3-319-22470-1} {\emph {\bibinfo {title}
  {Certified {{Reduced Basis Methods}} for {{Parametrized Partial Differential
  Equations}}}}},\ {{SpringerBriefs}} in {{Mathematics}}\ (\bibinfo
  {publisher} {{Springer International Publishing}},\ \bibinfo {address}
  {{Cham}},\ \bibinfo {year} {2016})\BibitemShut {NoStop}%
\bibitem [{\citenamefont {Quarteroni}\ \emph {et~al.}(2016)\citenamefont
  {Quarteroni}, \citenamefont {Manzoni},\ and\ \citenamefont
  {Negri}}]{quarteroniReducedBasisMethods2016}%
  \BibitemOpen
  \bibfield  {author} {\bibinfo {author} {\bibfnamefont {A.}~\bibnamefont
  {Quarteroni}}, \bibinfo {author} {\bibfnamefont {A.}~\bibnamefont
  {Manzoni}},\ and\ \bibinfo {author} {\bibfnamefont {F.}~\bibnamefont
  {Negri}},\ }\href {https://doi.org/10.1007/978-3-319-15431-2} {\emph
  {\bibinfo {title} {Reduced {{Basis Methods}} for {{Partial Differential
  Equations}}}}},\ \bibinfo {series} {{{UNITEXT}}}, Vol.~\bibinfo {volume}
  {92}\ (\bibinfo  {publisher} {{Springer International Publishing}},\ \bibinfo
  {address} {{Cham}},\ \bibinfo {year} {2016})\BibitemShut {NoStop}%
\bibitem [{\citenamefont {Herbst}\ \emph {et~al.}(2022)\citenamefont {Herbst},
  \citenamefont {Stamm}, \citenamefont {Wessel},\ and\ \citenamefont
  {Rizzi}}]{herbstSurrogateModelsQuantum2022}%
  \BibitemOpen
  \bibfield  {author} {\bibinfo {author} {\bibfnamefont {M.~F.}\ \bibnamefont
  {Herbst}}, \bibinfo {author} {\bibfnamefont {B.}~\bibnamefont {Stamm}},
  \bibinfo {author} {\bibfnamefont {S.}~\bibnamefont {Wessel}},\ and\ \bibinfo
  {author} {\bibfnamefont {M.}~\bibnamefont {Rizzi}},\ }\bibfield  {title}
  {\bibinfo {title} {Surrogate models for quantum spin systems based on
  reduced-order modeling},\ }\href
  {https://doi.org/10.1103/PhysRevE.105.045303} {\bibfield  {journal} {\bibinfo
   {journal} {Phys. Rev. E}\ }\textbf {\bibinfo {volume} {105}},\ \bibinfo
  {pages} {045303} (\bibinfo {year} {2022})}\BibitemShut {NoStop}%
\bibitem [{\citenamefont {Aktas}\ and\ \citenamefont
  {Moses}(1998)}]{aktasReducedBasisEigenvalue1998}%
  \BibitemOpen
  \bibfield  {author} {\bibinfo {author} {\bibfnamefont {E.}~\bibnamefont
  {Aktas}}\ and\ \bibinfo {author} {\bibfnamefont {F.}~\bibnamefont {Moses}},\
  }\bibfield  {title} {\bibinfo {title} {{Reduced Basis Eigenvalue Solutions}
  for {Damaged Structures}},\ }\href
  {https://doi.org/10.1080/08905459808945420} {\bibfield  {journal} {\bibinfo
  {journal} {Mech. Struct. Mach.}\ }\textbf {\bibinfo {volume} {26}},\ \bibinfo
  {pages} {63} (\bibinfo {year} {1998})}\BibitemShut {NoStop}%
\bibitem [{\citenamefont {Nair}\ \emph {et~al.}(1998)\citenamefont {Nair},
  \citenamefont {Keane},\ and\ \citenamefont
  {Langley}}]{nairImprovedFirstOrderApproximation1998}%
  \BibitemOpen
  \bibfield  {author} {\bibinfo {author} {\bibfnamefont {P.~B.}\ \bibnamefont
  {Nair}}, \bibinfo {author} {\bibfnamefont {A.~J.}\ \bibnamefont {Keane}},\
  and\ \bibinfo {author} {\bibfnamefont {R.~S.}\ \bibnamefont {Langley}},\
  }\bibfield  {title} {\bibinfo {title} {{Improved First-Order Approximation of
  Eigenvalues and Eigenvectors}},\ }\href {https://doi.org/10.2514/2.578}
  {\bibfield  {journal} {\bibinfo  {journal} {AIAA J.}\ }\textbf {\bibinfo
  {volume} {36}},\ \bibinfo {pages} {1721} (\bibinfo {year}
  {1998})}\BibitemShut {NoStop}%
\bibitem [{\citenamefont {Machiels}\ \emph {et~al.}(2000)\citenamefont
  {Machiels}, \citenamefont {Maday}, \citenamefont {Oliveira}, \citenamefont
  {Patera},\ and\ \citenamefont
  {Rovas}}]{machielsOutputBoundsReducedbasis2000}%
  \BibitemOpen
  \bibfield  {author} {\bibinfo {author} {\bibfnamefont {L.}~\bibnamefont
  {Machiels}}, \bibinfo {author} {\bibfnamefont {Y.}~\bibnamefont {Maday}},
  \bibinfo {author} {\bibfnamefont {I.~B.}\ \bibnamefont {Oliveira}}, \bibinfo
  {author} {\bibfnamefont {A.~T.}\ \bibnamefont {Patera}},\ and\ \bibinfo
  {author} {\bibfnamefont {D.~V.}\ \bibnamefont {Rovas}},\ }\bibfield  {title}
  {\bibinfo {title} {Output bounds for reduced-basis approximations of
  symmetric positive definite eigenvalue problems},\ }\href
  {https://doi.org/https://doi.org/10.1016/S0764-4442(00)00270-6} {\bibfield
  {journal} {\bibinfo  {journal} {Comptes Rendus de l'Académie des Sciences -
  Series I - Mathematics}\ }\textbf {\bibinfo {volume} {331}},\ \bibinfo
  {pages} {153} (\bibinfo {year} {2000})}\BibitemShut {NoStop}%
\bibitem [{\citenamefont {{Fumagalli, Ivan}}\ \emph {et~al.}(2016)\citenamefont
  {{Fumagalli, Ivan}}, \citenamefont {{Manzoni, Andrea}}, \citenamefont
  {{Parolini, Nicola}},\ and\ \citenamefont {{Verani,
  Marco}}}]{fumagalliReducedBasisApproximation2016}%
  \BibitemOpen
  \bibfield  {author} {\bibinfo {author} {\bibnamefont {{Fumagalli, Ivan}}},
  \bibinfo {author} {\bibnamefont {{Manzoni, Andrea}}}, \bibinfo {author}
  {\bibnamefont {{Parolini, Nicola}}},\ and\ \bibinfo {author} {\bibnamefont
  {{Verani, Marco}}},\ }\bibfield  {title} {\bibinfo {title} {Reduced basis
  approximation and a posteriori error estimates for parametrized elliptic
  eigenvalue problems},\ }\href {https://doi.org/10.1051/m2an/2016009}
  {\bibfield  {journal} {\bibinfo  {journal} {ESAIM: Math. Model. Numer.
  Anal.}\ }\textbf {\bibinfo {volume} {50}},\ \bibinfo {pages} {1857} (\bibinfo
  {year} {2016})}\BibitemShut {NoStop}%
\bibitem [{\citenamefont {{Horger, Thomas}}\ \emph {et~al.}(2017)\citenamefont
  {{Horger, Thomas}}, \citenamefont {{Wohlmuth, Barbara}},\ and\ \citenamefont
  {{Dickopf, Thomas}}}]{horgerSimultaneousReducedBasis2017}%
  \BibitemOpen
  \bibfield  {author} {\bibinfo {author} {\bibnamefont {{Horger, Thomas}}},
  \bibinfo {author} {\bibnamefont {{Wohlmuth, Barbara}}},\ and\ \bibinfo
  {author} {\bibnamefont {{Dickopf, Thomas}}},\ }\bibfield  {title} {\bibinfo
  {title} {Simultaneous reduced basis approximation of parameterized elliptic
  eigenvalue problems},\ }\href {https://doi.org/10.1051/m2an/2016025}
  {\bibfield  {journal} {\bibinfo  {journal} {ESAIM: Math. Model. Numer.
  Anal.}\ }\textbf {\bibinfo {volume} {51}},\ \bibinfo {pages} {443} (\bibinfo
  {year} {2017})}\BibitemShut {NoStop}%
\bibitem [{\citenamefont {Pla}\ and\ \citenamefont
  {Herrero}(2019)}]{plaReducedBasisMethod2019}%
  \BibitemOpen
  \bibfield  {author} {\bibinfo {author} {\bibfnamefont {F.}~\bibnamefont
  {Pla}}\ and\ \bibinfo {author} {\bibfnamefont {H.}~\bibnamefont {Herrero}},\
  }\bibfield  {title} {\bibinfo {title} {Reduced basis method applied to
  eigenvalue problems from convection},\ }\href
  {https://doi.org/10.1142/S0218127419500287} {\bibfield  {journal} {\bibinfo
  {journal} {Int. J. Bifurc. Chaos Appl. Sci. Eng.}\ }\textbf {\bibinfo
  {volume} {29}},\ \bibinfo {pages} {1950028} (\bibinfo {year}
  {2019})}\BibitemShut {NoStop}%
\bibitem [{\citenamefont {Pichi}\ \emph {et~al.}(2020)\citenamefont {Pichi},
  \citenamefont {Quaini},\ and\ \citenamefont {Rozza}}]{Pichi2020}%
  \BibitemOpen
  \bibfield  {author} {\bibinfo {author} {\bibfnamefont {F.}~\bibnamefont
  {Pichi}}, \bibinfo {author} {\bibfnamefont {A.}~\bibnamefont {Quaini}},\ and\
  \bibinfo {author} {\bibfnamefont {G.}~\bibnamefont {Rozza}},\ }\bibfield
  {title} {\bibinfo {title} {{A Reduced Order Modeling Technique to Study
  Bifurcating Phenomena: Application to the Gross--Pitaevskii Equation}},\
  }\href {https://doi.org/10.1137/20M1313106} {\bibfield  {journal} {\bibinfo
  {journal} {SIAM J. Sci. Comp.}\ }\textbf {\bibinfo {volume} {42}},\ \bibinfo
  {pages} {B1115} (\bibinfo {year} {2020})}\BibitemShut {NoStop}%
\bibitem [{\citenamefont {Ekstr\"om}\ and\ \citenamefont
  {Hagen}(2019)}]{Ekstroem2019}%
  \BibitemOpen
  \bibfield  {author} {\bibinfo {author} {\bibfnamefont {A.}~\bibnamefont
  {Ekstr\"om}}\ and\ \bibinfo {author} {\bibfnamefont {G.}~\bibnamefont
  {Hagen}},\ }\bibfield  {title} {\bibinfo {title} {{Global Sensitivity
  Analysis of Bulk Properties of an Atomic Nucleus}},\ }\href
  {https://doi.org/10.1103/PhysRevLett.123.252501} {\bibfield  {journal}
  {\bibinfo  {journal} {Phys. Rev. Lett.}\ }\textbf {\bibinfo {volume} {123}},\
  \bibinfo {pages} {252501} (\bibinfo {year} {2019})}\BibitemShut {NoStop}%
\bibitem [{\citenamefont {Demol}\ \emph {et~al.}(2020)\citenamefont {Demol},
  \citenamefont {Duguet}, \citenamefont {Ekstr\"om}, \citenamefont {Frosini},
  \citenamefont {Hebeler}, \citenamefont {K\"onig}, \citenamefont {Lee},
  \citenamefont {Schwenk}, \citenamefont {Som\`a},\ and\ \citenamefont
  {Tichai}}]{Demol2020}%
  \BibitemOpen
  \bibfield  {author} {\bibinfo {author} {\bibfnamefont {P.}~\bibnamefont
  {Demol}}, \bibinfo {author} {\bibfnamefont {T.}~\bibnamefont {Duguet}},
  \bibinfo {author} {\bibfnamefont {A.}~\bibnamefont {Ekstr\"om}}, \bibinfo
  {author} {\bibfnamefont {M.}~\bibnamefont {Frosini}}, \bibinfo {author}
  {\bibfnamefont {K.}~\bibnamefont {Hebeler}}, \bibinfo {author} {\bibfnamefont
  {S.}~\bibnamefont {K\"onig}}, \bibinfo {author} {\bibfnamefont
  {D.}~\bibnamefont {Lee}}, \bibinfo {author} {\bibfnamefont {A.}~\bibnamefont
  {Schwenk}}, \bibinfo {author} {\bibfnamefont {V.}~\bibnamefont {Som\`a}},\
  and\ \bibinfo {author} {\bibfnamefont {A.}~\bibnamefont {Tichai}},\
  }\bibfield  {title} {\bibinfo {title} {Improved many-body expansions from
  eigenvector continuation},\ }\href
  {https://doi.org/10.1103/PhysRevC.101.041302} {\bibfield  {journal} {\bibinfo
   {journal} {Phys. Rev. C}\ }\textbf {\bibinfo {volume} {101}},\ \bibinfo
  {pages} {041302} (\bibinfo {year} {2020})}\BibitemShut {NoStop}%
\bibitem [{\citenamefont {K\"onig}\ \emph {et~al.}(2020)\citenamefont
  {K\"onig}, \citenamefont {Ekstr\"om}, \citenamefont {Hebeler}, \citenamefont
  {Lee},\ and\ \citenamefont {Schwenk}}]{Koenig2020}%
  \BibitemOpen
  \bibfield  {author} {\bibinfo {author} {\bibfnamefont {S.}~\bibnamefont
  {K\"onig}}, \bibinfo {author} {\bibfnamefont {A.}~\bibnamefont {Ekstr\"om}},
  \bibinfo {author} {\bibfnamefont {K.}~\bibnamefont {Hebeler}}, \bibinfo
  {author} {\bibfnamefont {D.}~\bibnamefont {Lee}},\ and\ \bibinfo {author}
  {\bibfnamefont {A.}~\bibnamefont {Schwenk}},\ }\bibfield  {title} {\bibinfo
  {title} {Eigenvector continuation as an efficient and accurate emulator for
  uncertainty quantification},\ }\href
  {https://doi.org/https://doi.org/10.1016/j.physletb.2020.135814} {\bibfield
  {journal} {\bibinfo  {journal} {Phys. Lett. B}\ }\textbf {\bibinfo {volume}
  {810}},\ \bibinfo {pages} {135814} (\bibinfo {year} {2020})}\BibitemShut
  {NoStop}%
\bibitem [{\citenamefont {Furnstahl}\ \emph {et~al.}(2020)\citenamefont
  {Furnstahl}, \citenamefont {Garcia}, \citenamefont {Millican},\ and\
  \citenamefont {Zhang}}]{Furnstahl2020}%
  \BibitemOpen
  \bibfield  {author} {\bibinfo {author} {\bibfnamefont {R.}~\bibnamefont
  {Furnstahl}}, \bibinfo {author} {\bibfnamefont {A.}~\bibnamefont {Garcia}},
  \bibinfo {author} {\bibfnamefont {P.}~\bibnamefont {Millican}},\ and\
  \bibinfo {author} {\bibfnamefont {X.}~\bibnamefont {Zhang}},\ }\bibfield
  {title} {\bibinfo {title} {Efficient emulators for scattering using
  eigenvector continuation},\ }\href
  {https://doi.org/https://doi.org/10.1016/j.physletb.2020.135719} {\bibfield
  {journal} {\bibinfo  {journal} {Phys. Lett. B}\ }\textbf {\bibinfo {volume}
  {809}},\ \bibinfo {pages} {135719} (\bibinfo {year} {2020})}\BibitemShut
  {NoStop}%
\bibitem [{\citenamefont {Drischler}\ \emph {et~al.}(2021)\citenamefont
  {Drischler}, \citenamefont {Quinonez}, \citenamefont {Giuliani},
  \citenamefont {Lovell},\ and\ \citenamefont {Nunes}}]{Drischler2021}%
  \BibitemOpen
  \bibfield  {author} {\bibinfo {author} {\bibfnamefont {C.}~\bibnamefont
  {Drischler}}, \bibinfo {author} {\bibfnamefont {M.}~\bibnamefont {Quinonez}},
  \bibinfo {author} {\bibfnamefont {P.}~\bibnamefont {Giuliani}}, \bibinfo
  {author} {\bibfnamefont {A.}~\bibnamefont {Lovell}},\ and\ \bibinfo {author}
  {\bibfnamefont {F.}~\bibnamefont {Nunes}},\ }\bibfield  {title} {\bibinfo
  {title} {Toward emulating nuclear reactions using eigenvector continuation},\
  }\href {https://doi.org/https://doi.org/10.1016/j.physletb.2021.136777}
  {\bibfield  {journal} {\bibinfo  {journal} {Phys. Lett. B}\ }\textbf
  {\bibinfo {volume} {823}},\ \bibinfo {pages} {136777} (\bibinfo {year}
  {2021})}\BibitemShut {NoStop}%
\bibitem [{\citenamefont {Bonilla}\ \emph {et~al.}(2022)\citenamefont
  {Bonilla}, \citenamefont {Giuliani}, \citenamefont {Godbey},\ and\
  \citenamefont {Lee}}]{Bonilla2022}%
  \BibitemOpen
  \bibfield  {author} {\bibinfo {author} {\bibfnamefont {E.}~\bibnamefont
  {Bonilla}}, \bibinfo {author} {\bibfnamefont {P.}~\bibnamefont {Giuliani}},
  \bibinfo {author} {\bibfnamefont {K.}~\bibnamefont {Godbey}},\ and\ \bibinfo
  {author} {\bibfnamefont {D.}~\bibnamefont {Lee}},\ }\bibfield  {title}
  {\bibinfo {title} {Training and projecting: A reduced basis method emulator
  for many-body physics},\ }\href {https://doi.org/10.1103/PhysRevC.106.054322}
  {\bibfield  {journal} {\bibinfo  {journal} {Phys. Rev. C}\ }\textbf {\bibinfo
  {volume} {106}},\ \bibinfo {pages} {054322} (\bibinfo {year}
  {2022})}\BibitemShut {NoStop}%
\bibitem [{\citenamefont {Frame}\ \emph {et~al.}(2018)\citenamefont {Frame},
  \citenamefont {He}, \citenamefont {Ipsen}, \citenamefont {Lee}, \citenamefont
  {Lee},\ and\ \citenamefont
  {Rrapaj}}]{frameEigenvectorContinuationSubspace2018}%
  \BibitemOpen
  \bibfield  {author} {\bibinfo {author} {\bibfnamefont {D.}~\bibnamefont
  {Frame}}, \bibinfo {author} {\bibfnamefont {R.}~\bibnamefont {He}}, \bibinfo
  {author} {\bibfnamefont {I.}~\bibnamefont {Ipsen}}, \bibinfo {author}
  {\bibfnamefont {D.}~\bibnamefont {Lee}}, \bibinfo {author} {\bibfnamefont
  {D.}~\bibnamefont {Lee}},\ and\ \bibinfo {author} {\bibfnamefont
  {E.}~\bibnamefont {Rrapaj}},\ }\bibfield  {title} {\bibinfo {title}
  {Eigenvector {{Continuation}} with {{Subspace Learning}}},\ }\href
  {https://doi.org/10.1103/PhysRevLett.121.032501} {\bibfield  {journal}
  {\bibinfo  {journal} {Phys. Rev. Lett.}\ }\textbf {\bibinfo {volume} {121}},\
  \bibinfo {pages} {032501} (\bibinfo {year} {2018})}\BibitemShut {NoStop}%
\bibitem [{\citenamefont {Sarkar}\ and\ \citenamefont
  {Lee}(2021)}]{Sarkar2021}%
  \BibitemOpen
  \bibfield  {author} {\bibinfo {author} {\bibfnamefont {A.}~\bibnamefont
  {Sarkar}}\ and\ \bibinfo {author} {\bibfnamefont {D.}~\bibnamefont {Lee}},\
  }\bibfield  {title} {\bibinfo {title} {{Convergence of Eigenvector
  Continuation}},\ }\href {https://doi.org/10.1103/PhysRevLett.126.032501}
  {\bibfield  {journal} {\bibinfo  {journal} {Phys. Rev. Lett.}\ }\textbf
  {\bibinfo {volume} {126}},\ \bibinfo {pages} {032501} (\bibinfo {year}
  {2021})}\BibitemShut {NoStop}%
\bibitem [{\citenamefont {Sarkar}\ and\ \citenamefont
  {Lee}(2022)}]{sarkarSelflearningEmulatorsEigenvector2022}%
  \BibitemOpen
  \bibfield  {author} {\bibinfo {author} {\bibfnamefont {A.}~\bibnamefont
  {Sarkar}}\ and\ \bibinfo {author} {\bibfnamefont {D.}~\bibnamefont {Lee}},\
  }\bibfield  {title} {\bibinfo {title} {Self-learning emulators and
  eigenvector continuation},\ }\href
  {https://doi.org/10.1103/PhysRevResearch.4.023214} {\bibfield  {journal}
  {\bibinfo  {journal} {Phys. Rev. Res.}\ }\textbf {\bibinfo {volume} {4}},\
  \bibinfo {pages} {023214} (\bibinfo {year} {2022})}\BibitemShut {NoStop}%
\bibitem [{\citenamefont {Melendez}\ \emph {et~al.}(2022)\citenamefont
  {Melendez}, \citenamefont {Drischler}, \citenamefont {Furnstahl},
  \citenamefont {Garcia},\ and\ \citenamefont {Zhang}}]{Melendez2022}%
  \BibitemOpen
  \bibfield  {author} {\bibinfo {author} {\bibfnamefont {J.~A.}\ \bibnamefont
  {Melendez}}, \bibinfo {author} {\bibfnamefont {C.}~\bibnamefont {Drischler}},
  \bibinfo {author} {\bibfnamefont {R.~J.}\ \bibnamefont {Furnstahl}}, \bibinfo
  {author} {\bibfnamefont {A.~J.}\ \bibnamefont {Garcia}},\ and\ \bibinfo
  {author} {\bibfnamefont {X.}~\bibnamefont {Zhang}},\ }\bibfield  {title}
  {\bibinfo {title} {Model reduction methods for nuclear emulators},\ }\href
  {https://doi.org/10.1088/1361-6471/ac83dd} {\bibfield  {journal} {\bibinfo
  {journal} {J. Phys. G}\ }\textbf {\bibinfo {volume} {49}},\ \bibinfo {pages}
  {102001} (\bibinfo {year} {2022})}\BibitemShut {NoStop}%
\bibitem [{\citenamefont {Francis}\ \emph {et~al.}(2022)\citenamefont
  {Francis}, \citenamefont {Agrawal}, \citenamefont {Howard}, \citenamefont
  {K\"okc\"u},\ and\ \citenamefont
  {Kemper}}]{francisSubspaceDiagonalizationQuantum2022}%
  \BibitemOpen
  \bibfield  {author} {\bibinfo {author} {\bibfnamefont {A.}~\bibnamefont
  {Francis}}, \bibinfo {author} {\bibfnamefont {A.~A.}\ \bibnamefont
  {Agrawal}}, \bibinfo {author} {\bibfnamefont {J.~H.}\ \bibnamefont {Howard}},
  \bibinfo {author} {\bibfnamefont {E.}~\bibnamefont {K\"okc\"u}},\ and\
  \bibinfo {author} {\bibfnamefont {A.~F.}\ \bibnamefont {Kemper}},\
  }\href@noop {} {\bibinfo {title} {Subspace diagonalization on quantum
  computers using eigenvector continuation}} (\bibinfo {year} {2022}),\ \Eprint
  {https://arxiv.org/abs/2209.10571} {arXiv:2209.10571} \BibitemShut {NoStop}%
\bibitem [{\citenamefont {Baran}\ and\ \citenamefont
  {Nichita}(2023)}]{baranReducedBasisEmulation2022}%
  \BibitemOpen
  \bibfield  {author} {\bibinfo {author} {\bibfnamefont {V.~V.}\ \bibnamefont
  {Baran}}\ and\ \bibinfo {author} {\bibfnamefont {D.~R.}\ \bibnamefont
  {Nichita}},\ }\bibfield  {title} {\bibinfo {title} {Reduced basis emulation
  of pairing in finite systems},\ }\href
  {https://doi.org/10.1103/PhysRevB.107.144503} {\bibfield  {journal} {\bibinfo
   {journal} {Phys. Rev. B}\ }\textbf {\bibinfo {volume} {107}},\ \bibinfo
  {pages} {144503} (\bibinfo {year} {2023})}\BibitemShut {NoStop}%
\bibitem [{\citenamefont {Franzke}\ \emph {et~al.}(2023)\citenamefont
  {Franzke}, \citenamefont {Tichai}, \citenamefont {Hebeler},\ and\
  \citenamefont {Schwenk}}]{CompanysFranzke2023}%
  \BibitemOpen
  \bibfield  {author} {\bibinfo {author} {\bibfnamefont {M.~C.}\ \bibnamefont
  {Franzke}}, \bibinfo {author} {\bibfnamefont {A.}~\bibnamefont {Tichai}},
  \bibinfo {author} {\bibfnamefont {K.}~\bibnamefont {Hebeler}},\ and\ \bibinfo
  {author} {\bibfnamefont {A.}~\bibnamefont {Schwenk}},\ }\href@noop {}
  {\bibinfo {title} {{Eigenvector continuation for the pairing Hamiltonian}}}
  (\bibinfo {year} {2023}),\ \Eprint {https://arxiv.org/abs/2302.08373}
  {arXiv:2302.08373} \BibitemShut {NoStop}%
\bibitem [{\citenamefont {Mejuto-Zaera}\ and\ \citenamefont
  {Kemper}(2023)}]{MejutoZaera2023}%
  \BibitemOpen
  \bibfield  {author} {\bibinfo {author} {\bibfnamefont {C.}~\bibnamefont
  {Mejuto-Zaera}}\ and\ \bibinfo {author} {\bibfnamefont {A.~F.}\ \bibnamefont
  {Kemper}},\ }\href@noop {} {\bibinfo {title} {{Quantum Eigenvector
  Continuation for Chemistry Applications}}} (\bibinfo {year} {2023}),\ \Eprint
  {https://arxiv.org/abs/2305.00060} {arXiv:2305.00060} \BibitemShut {NoStop}%
\bibitem [{\citenamefont {White}(1992)}]{White1992}%
  \BibitemOpen
  \bibfield  {author} {\bibinfo {author} {\bibfnamefont {S.~R.}\ \bibnamefont
  {White}},\ }\bibfield  {title} {\bibinfo {title} {Density matrix formulation
  for quantum renormalization groups},\ }\href
  {https://doi.org/10.1103/PhysRevLett.69.2863} {\bibfield  {journal} {\bibinfo
   {journal} {Phys. Rev. Lett.}\ }\textbf {\bibinfo {volume} {69}},\ \bibinfo
  {pages} {2863} (\bibinfo {year} {1992})}\BibitemShut {NoStop}%
\bibitem [{\citenamefont {White}(1993)}]{White1993}%
  \BibitemOpen
  \bibfield  {author} {\bibinfo {author} {\bibfnamefont {S.~R.}\ \bibnamefont
  {White}},\ }\bibfield  {title} {\bibinfo {title} {Density-matrix algorithms
  for quantum renormalization groups},\ }\href
  {https://doi.org/10.1103/PhysRevB.48.10345} {\bibfield  {journal} {\bibinfo
  {journal} {Phys. Rev. B}\ }\textbf {\bibinfo {volume} {48}},\ \bibinfo
  {pages} {10345} (\bibinfo {year} {1993})}\BibitemShut {NoStop}%
\bibitem [{\citenamefont {Fannes}\ \emph {et~al.}(1992)\citenamefont {Fannes},
  \citenamefont {Nachtergaele},\ and\ \citenamefont {Werner}}]{Fannes1992}%
  \BibitemOpen
  \bibfield  {author} {\bibinfo {author} {\bibfnamefont {M.}~\bibnamefont
  {Fannes}}, \bibinfo {author} {\bibfnamefont {B.}~\bibnamefont
  {Nachtergaele}},\ and\ \bibinfo {author} {\bibfnamefont {R.~F.}\ \bibnamefont
  {Werner}},\ }\bibfield  {title} {\bibinfo {title} {{Finitely Correlated
  States on Quantum Spin Chains}},\ }\href {https://doi.org/10.1007/BF02099178}
  {\bibfield  {journal} {\bibinfo  {journal} {Commun. Math. Phys.}\ }\textbf
  {\bibinfo {volume} {144}},\ \bibinfo {pages} {443} (\bibinfo {year}
  {1992})}\BibitemShut {NoStop}%
\bibitem [{\citenamefont {Kl\"umper}\ \emph {et~al.}(1993)\citenamefont
  {Kl\"umper}, \citenamefont {Schadschneider},\ and\ \citenamefont
  {Zittartz}}]{Kluemper1993}%
  \BibitemOpen
  \bibfield  {author} {\bibinfo {author} {\bibfnamefont {A.}~\bibnamefont
  {Kl\"umper}}, \bibinfo {author} {\bibfnamefont {A.}~\bibnamefont
  {Schadschneider}},\ and\ \bibinfo {author} {\bibfnamefont {J.}~\bibnamefont
  {Zittartz}},\ }\bibfield  {title} {\bibinfo {title} {{Matrix Product Ground
  States for One-Dimensional Spin-1 Quantum Antiferromagnets}},\ }\href
  {https://doi.org/10.1209/0295-5075/24/4/010} {\bibfield  {journal} {\bibinfo
  {journal} {EPL}\ }\textbf {\bibinfo {volume} {24}},\ \bibinfo {pages} {293}
  (\bibinfo {year} {1993})}\BibitemShut {NoStop}%
\bibitem [{\citenamefont {\"Ostlund}\ and\ \citenamefont
  {Rommer}(1995)}]{Ostlund1995}%
  \BibitemOpen
  \bibfield  {author} {\bibinfo {author} {\bibfnamefont {S.}~\bibnamefont
  {\"Ostlund}}\ and\ \bibinfo {author} {\bibfnamefont {S.}~\bibnamefont
  {Rommer}},\ }\bibfield  {title} {\bibinfo {title} {{Thermodynamic Limit of
  Density Matrix Renormalization}},\ }\href
  {https://doi.org/10.1103/PhysRevLett.75.3537} {\bibfield  {journal} {\bibinfo
   {journal} {Phys. Rev. Lett.}\ }\textbf {\bibinfo {volume} {75}},\ \bibinfo
  {pages} {3537} (\bibinfo {year} {1995})}\BibitemShut {NoStop}%
\bibitem [{\citenamefont
  {Schollw\"ock}(2011)}]{schollwoeckDensitymatrixRenormalizationGroup2011}%
  \BibitemOpen
  \bibfield  {author} {\bibinfo {author} {\bibfnamefont {U.}~\bibnamefont
  {Schollw\"ock}},\ }\bibfield  {title} {\bibinfo {title} {The density-matrix
  renormalization group in the age of matrix product states},\ }\href
  {https://doi.org/https://doi.org/10.1016/j.aop.2010.09.012} {\bibfield
  {journal} {\bibinfo  {journal} {Ann. Phys.}\ }\textbf {\bibinfo {volume}
  {326}},\ \bibinfo {pages} {96} (\bibinfo {year} {2011})}\BibitemShut
  {NoStop}%
\bibitem [{\citenamefont {Verstraete}\ \emph {et~al.}(2008)\citenamefont
  {Verstraete}, \citenamefont {Murg},\ and\ \citenamefont
  {Cirac}}]{Verstraete2008}%
  \BibitemOpen
  \bibfield  {author} {\bibinfo {author} {\bibfnamefont {F.}~\bibnamefont
  {Verstraete}}, \bibinfo {author} {\bibfnamefont {V.}~\bibnamefont {Murg}},\
  and\ \bibinfo {author} {\bibfnamefont {J.}~\bibnamefont {Cirac}},\ }\bibfield
   {title} {\bibinfo {title} {Matrix product states, projected entangled pair
  states, and variational renormalization group methods for quantum spin
  systems},\ }\href {https://doi.org/10.1080/14789940801912366} {\bibfield
  {journal} {\bibinfo  {journal} {Adv. Phys.}\ }\textbf {\bibinfo {volume}
  {57}},\ \bibinfo {pages} {143} (\bibinfo {year} {2008})}\BibitemShut
  {NoStop}%
\bibitem [{\citenamefont {Orús}(2014)}]{orusPracticalIntroductionTensor2014}%
  \BibitemOpen
  \bibfield  {author} {\bibinfo {author} {\bibfnamefont {R.}~\bibnamefont
  {Orús}},\ }\bibfield  {title} {\bibinfo {title} {A practical introduction to
  tensor networks: {Matrix} product states and projected entangled pair
  states},\ }\href {https://doi.org/https://doi.org/10.1016/j.aop.2014.06.013}
  {\bibfield  {journal} {\bibinfo  {journal} {Ann. Phys. (N. Y.)}\ }\textbf
  {\bibinfo {volume} {349}},\ \bibinfo {pages} {117} (\bibinfo {year}
  {2014})}\BibitemShut {NoStop}%
\bibitem [{Git()}]{GitHubRepo}%
  \BibitemOpen
  \href@noop {} {}\bibinfo {howpublished}
  {\url{https://github.com/mfherbst/ReducedBasis.jl}}\BibitemShut {NoStop}%
\bibitem [{\citenamefont {L\"owdin}(1950)}]{Loewdin1950}%
  \BibitemOpen
  \bibfield  {author} {\bibinfo {author} {\bibfnamefont {P.}~\bibnamefont
  {L\"owdin}},\ }\bibfield  {title} {\bibinfo {title} {{On the
  Non‐Orthogonality Problem Connected with the Use of Atomic Wave Functions
  in the Theory of Molecules and Crystals}},\ }\href
  {https://doi.org/10.1063/1.1747632} {\bibfield  {journal} {\bibinfo
  {journal} {J. Chem. Phys.}\ }\textbf {\bibinfo {volume} {18}},\ \bibinfo
  {pages} {365} (\bibinfo {year} {1950})}\BibitemShut {NoStop}%
\bibitem [{\citenamefont {Verstraete}\ and\ \citenamefont
  {Cirac}(2004)}]{Verstraete2004}%
  \BibitemOpen
  \bibfield  {author} {\bibinfo {author} {\bibfnamefont {F.}~\bibnamefont
  {Verstraete}}\ and\ \bibinfo {author} {\bibfnamefont {J.~I.}\ \bibnamefont
  {Cirac}},\ }\href@noop {} {\bibinfo {title} {Renormalization algorithms for
  {Quantum-Many Body Systems} in two and higher dimensions}} (\bibinfo {year}
  {2004}),\ \Eprint {https://arxiv.org/abs/cond-mat/0407066}
  {arXiv:cond-mat/0407066} \BibitemShut {NoStop}%
\bibitem [{\citenamefont {Shi}\ \emph {et~al.}(2006)\citenamefont {Shi},
  \citenamefont {Duan},\ and\ \citenamefont {Vidal}}]{Shi2006}%
  \BibitemOpen
  \bibfield  {author} {\bibinfo {author} {\bibfnamefont {Y.-Y.}\ \bibnamefont
  {Shi}}, \bibinfo {author} {\bibfnamefont {L.-M.}\ \bibnamefont {Duan}},\ and\
  \bibinfo {author} {\bibfnamefont {G.}~\bibnamefont {Vidal}},\ }\bibfield
  {title} {\bibinfo {title} {Classical simulation of quantum many-body systems
  with a tree tensor network},\ }\href
  {https://doi.org/10.1103/PhysRevA.74.022320} {\bibfield  {journal} {\bibinfo
  {journal} {Phys. Rev. A}\ }\textbf {\bibinfo {volume} {74}},\ \bibinfo
  {pages} {022320} (\bibinfo {year} {2006})}\BibitemShut {NoStop}%
\bibitem [{\citenamefont {Gerster}\ \emph {et~al.}(2014)\citenamefont
  {Gerster}, \citenamefont {Silvi}, \citenamefont {Rizzi}, \citenamefont
  {Fazio}, \citenamefont {Calarco},\ and\ \citenamefont
  {Montangero}}]{Gerster2014}%
  \BibitemOpen
  \bibfield  {author} {\bibinfo {author} {\bibfnamefont {M.}~\bibnamefont
  {Gerster}}, \bibinfo {author} {\bibfnamefont {P.}~\bibnamefont {Silvi}},
  \bibinfo {author} {\bibfnamefont {M.}~\bibnamefont {Rizzi}}, \bibinfo
  {author} {\bibfnamefont {R.}~\bibnamefont {Fazio}}, \bibinfo {author}
  {\bibfnamefont {T.}~\bibnamefont {Calarco}},\ and\ \bibinfo {author}
  {\bibfnamefont {S.}~\bibnamefont {Montangero}},\ }\bibfield  {title}
  {\bibinfo {title} {Unconstrained tree tensor network: An adaptive gauge
  picture for enhanced performance},\ }\href
  {https://doi.org/10.1103/PhysRevB.90.125154} {\bibfield  {journal} {\bibinfo
  {journal} {Phys. Rev. B}\ }\textbf {\bibinfo {volume} {90}},\ \bibinfo
  {pages} {125154} (\bibinfo {year} {2014})}\BibitemShut {NoStop}%
\bibitem [{\citenamefont {Gerster}\ \emph {et~al.}(2017)\citenamefont
  {Gerster}, \citenamefont {Rizzi}, \citenamefont {Silvi}, \citenamefont
  {Dalmonte},\ and\ \citenamefont {Montangero}}]{Gerster2017}%
  \BibitemOpen
  \bibfield  {author} {\bibinfo {author} {\bibfnamefont {M.}~\bibnamefont
  {Gerster}}, \bibinfo {author} {\bibfnamefont {M.}~\bibnamefont {Rizzi}},
  \bibinfo {author} {\bibfnamefont {P.}~\bibnamefont {Silvi}}, \bibinfo
  {author} {\bibfnamefont {M.}~\bibnamefont {Dalmonte}},\ and\ \bibinfo
  {author} {\bibfnamefont {S.}~\bibnamefont {Montangero}},\ }\bibfield  {title}
  {\bibinfo {title} {Fractional quantum hall effect in the interacting
  {Hofstadter} model via tensor networks},\ }\href
  {https://doi.org/10.1103/PhysRevB.96.195123} {\bibfield  {journal} {\bibinfo
  {journal} {Phys. Rev. B}\ }\textbf {\bibinfo {volume} {96}},\ \bibinfo
  {pages} {195123} (\bibinfo {year} {2017})}\BibitemShut {NoStop}%
\bibitem [{\citenamefont {Carleo}\ and\ \citenamefont
  {Troyer}(2017)}]{Caleo2017}%
  \BibitemOpen
  \bibfield  {author} {\bibinfo {author} {\bibfnamefont {G.}~\bibnamefont
  {Carleo}}\ and\ \bibinfo {author} {\bibfnamefont {M.}~\bibnamefont
  {Troyer}},\ }\bibfield  {title} {\bibinfo {title} {Solving the quantum
  many-body problem with artificial neural networks},\ }\href
  {https://doi.org/10.1126/science.aag2302} {\bibfield  {journal} {\bibinfo
  {journal} {Science}\ }\textbf {\bibinfo {volume} {355}},\ \bibinfo {pages}
  {602} (\bibinfo {year} {2017})}\BibitemShut {NoStop}%
\bibitem [{\citenamefont {Gubernatis}\ \emph {et~al.}(2016)\citenamefont
  {Gubernatis}, \citenamefont {Kawashima},\ and\ \citenamefont
  {Werner}}]{Gubernatis2016}%
  \BibitemOpen
  \bibfield  {author} {\bibinfo {author} {\bibfnamefont {J.}~\bibnamefont
  {Gubernatis}}, \bibinfo {author} {\bibfnamefont {N.}~\bibnamefont
  {Kawashima}},\ and\ \bibinfo {author} {\bibfnamefont {P.}~\bibnamefont
  {Werner}},\ }\href {https://doi.org/10.1017/CBO9780511902581} {\emph
  {\bibinfo {title} {Quantum Monte Carlo Methods: Algorithms for Lattice
  Models}}}\ (\bibinfo  {publisher} {Cambridge University Press},\ \bibinfo
  {year} {2016})\BibitemShut {NoStop}%
\bibitem [{\citenamefont {Bezanson}\ \emph {et~al.}(2017)\citenamefont
  {Bezanson}, \citenamefont {Edelman}, \citenamefont {Karpinski},\ and\
  \citenamefont {Shah}}]{bezansonJuliaFreshApproach2017}%
  \BibitemOpen
  \bibfield  {author} {\bibinfo {author} {\bibfnamefont {J.}~\bibnamefont
  {Bezanson}}, \bibinfo {author} {\bibfnamefont {A.}~\bibnamefont {Edelman}},
  \bibinfo {author} {\bibfnamefont {S.}~\bibnamefont {Karpinski}},\ and\
  \bibinfo {author} {\bibfnamefont {V.~B.}\ \bibnamefont {Shah}},\ }\bibfield
  {title} {\bibinfo {title} {{Julia: A Fresh Approach} to {Numerical
  Computing}},\ }\href {https://doi.org/10.1137/141000671} {\bibfield
  {journal} {\bibinfo  {journal} {SIAM Rev.}\ }\textbf {\bibinfo {volume}
  {59}},\ \bibinfo {pages} {65} (\bibinfo {year} {2017})}\BibitemShut {NoStop}%
\bibitem [{\citenamefont {Fishman}\ \emph
  {et~al.}(2022{\natexlab{a}})\citenamefont {Fishman}, \citenamefont {White},\
  and\ \citenamefont {Stoudenmire}}]{fishmanITensorSoftwareLibrary2022}%
  \BibitemOpen
  \bibfield  {author} {\bibinfo {author} {\bibfnamefont {M.}~\bibnamefont
  {Fishman}}, \bibinfo {author} {\bibfnamefont {S.~R.}\ \bibnamefont {White}},\
  and\ \bibinfo {author} {\bibfnamefont {E.~M.}\ \bibnamefont {Stoudenmire}},\
  }\bibfield  {title} {\bibinfo {title} {{The ITensor Software Library for
  Tensor Network Calculations}},\ }\href
  {https://doi.org/10.21468/SciPostPhysCodeb.4} {\bibfield  {journal} {\bibinfo
   {journal} {SciPost Phys. Codebases}\ ,\ \bibinfo {pages} {4}} (\bibinfo
  {year} {2022}{\natexlab{a}})}\BibitemShut {NoStop}%
\bibitem [{\citenamefont {Fishman}\ \emph
  {et~al.}(2022{\natexlab{b}})\citenamefont {Fishman}, \citenamefont {White},\
  and\ \citenamefont {Stoudenmire}}]{fishmanITensor03Release2022}%
  \BibitemOpen
  \bibfield  {author} {\bibinfo {author} {\bibfnamefont {M.}~\bibnamefont
  {Fishman}}, \bibinfo {author} {\bibfnamefont {S.~R.}\ \bibnamefont {White}},\
  and\ \bibinfo {author} {\bibfnamefont {E.~M.}\ \bibnamefont {Stoudenmire}},\
  }\bibfield  {title} {\bibinfo {title} {{Codebase release 0.3 for ITensor}},\
  }\href {https://doi.org/10.21468/SciPostPhysCodeb.4-r0.3} {\bibfield
  {journal} {\bibinfo  {journal} {SciPost Phys. Codebases}\ ,\ \bibinfo {pages}
  {4}} (\bibinfo {year} {2022}{\natexlab{b}})}\BibitemShut {NoStop}%
\bibitem [{\citenamefont {Tzeng}(2012)}]{Tzeng2012}%
  \BibitemOpen
  \bibfield  {author} {\bibinfo {author} {\bibfnamefont {Y.-C.}\ \bibnamefont
  {Tzeng}},\ }\bibfield  {title} {\bibinfo {title} {Parity quantum numbers in
  the density matrix renormalization group},\ }\href
  {https://doi.org/10.1103/PhysRevB.86.024403} {\bibfield  {journal} {\bibinfo
  {journal} {Phys. Rev. B}\ }\textbf {\bibinfo {volume} {86}},\ \bibinfo
  {pages} {024403} (\bibinfo {year} {2012})}\BibitemShut {NoStop}%
\bibitem [{\citenamefont {Affleck}\ \emph {et~al.}(1987)\citenamefont
  {Affleck}, \citenamefont {Kennedy}, \citenamefont {Lieb},\ and\ \citenamefont
  {Tasaki}}]{Affleck1987}%
  \BibitemOpen
  \bibfield  {author} {\bibinfo {author} {\bibfnamefont {I.}~\bibnamefont
  {Affleck}}, \bibinfo {author} {\bibfnamefont {T.}~\bibnamefont {Kennedy}},
  \bibinfo {author} {\bibfnamefont {E.~H.}\ \bibnamefont {Lieb}},\ and\
  \bibinfo {author} {\bibfnamefont {H.}~\bibnamefont {Tasaki}},\ }\bibfield
  {title} {\bibinfo {title} {Rigorous results on valence-bond ground states in
  antiferromagnets},\ }\href {https://doi.org/10.1103/PhysRevLett.59.799}
  {\bibfield  {journal} {\bibinfo  {journal} {Phys. Rev. Lett.}\ }\textbf
  {\bibinfo {volume} {59}},\ \bibinfo {pages} {799} (\bibinfo {year}
  {1987})}\BibitemShut {NoStop}%
\bibitem [{Note1()}]{Note1}%
  \BibitemOpen
  \bibinfo {note} {Note that in the ITensor convention, the singular value
  cutoffs are defined via the squared Frobenius norm, such that the ITensor
  cutoffs correspond to $\protect \mathtt {cut}_\sigma ^2$.}\BibitemShut
  {Stop}%
\bibitem [{\citenamefont {Vidal}(2007)}]{Vidal2007}%
  \BibitemOpen
  \bibfield  {author} {\bibinfo {author} {\bibfnamefont {G.}~\bibnamefont
  {Vidal}},\ }\bibfield  {title} {\bibinfo {title} {{Classical Simulation of
  Infinite-Size Quantum Lattice Systems in One Spatial Dimension}},\ }\href
  {https://doi.org/10.1103/PhysRevLett.98.070201} {\bibfield  {journal}
  {\bibinfo  {journal} {Phys. Rev. Lett.}\ }\textbf {\bibinfo {volume} {98}},\
  \bibinfo {pages} {070201} (\bibinfo {year} {2007})}\BibitemShut {NoStop}%
\bibitem [{\citenamefont {McCulloch}(2008)}]{McCulloch2008}%
  \BibitemOpen
  \bibfield  {author} {\bibinfo {author} {\bibfnamefont {I.~P.}\ \bibnamefont
  {McCulloch}},\ }\href@noop {} {\bibinfo {title} {Infinite size density matrix
  renormalization group, revisited}} (\bibinfo {year} {2008}),\ \Eprint
  {https://arxiv.org/abs/0804.2509} {arXiv:0804.2509} \BibitemShut {NoStop}%
\bibitem [{\citenamefont {Zauner-Stauber}\ \emph {et~al.}(2018)\citenamefont
  {Zauner-Stauber}, \citenamefont {Vanderstraeten}, \citenamefont {Fishman},
  \citenamefont {Verstraete},\ and\ \citenamefont
  {Haegeman}}]{ZaunerStauber2018}%
  \BibitemOpen
  \bibfield  {author} {\bibinfo {author} {\bibfnamefont {V.}~\bibnamefont
  {Zauner-Stauber}}, \bibinfo {author} {\bibfnamefont {L.}~\bibnamefont
  {Vanderstraeten}}, \bibinfo {author} {\bibfnamefont {M.~T.}\ \bibnamefont
  {Fishman}}, \bibinfo {author} {\bibfnamefont {F.}~\bibnamefont
  {Verstraete}},\ and\ \bibinfo {author} {\bibfnamefont {J.}~\bibnamefont
  {Haegeman}},\ }\bibfield  {title} {\bibinfo {title} {Variational optimization
  algorithms for uniform matrix product states},\ }\href
  {https://doi.org/10.1103/PhysRevB.97.045145} {\bibfield  {journal} {\bibinfo
  {journal} {Phys. Rev. B}\ }\textbf {\bibinfo {volume} {97}},\ \bibinfo
  {pages} {045145} (\bibinfo {year} {2018})}\BibitemShut {NoStop}%
\bibitem [{\citenamefont {Binev}\ \emph {et~al.}(2011)\citenamefont {Binev},
  \citenamefont {Cohen}, \citenamefont {Dahmen}, \citenamefont {DeVore},
  \citenamefont {Petrova},\ and\ \citenamefont {Wojtaszczyk}}]{Binev2011}%
  \BibitemOpen
  \bibfield  {author} {\bibinfo {author} {\bibfnamefont {P.}~\bibnamefont
  {Binev}}, \bibinfo {author} {\bibfnamefont {A.}~\bibnamefont {Cohen}},
  \bibinfo {author} {\bibfnamefont {W.}~\bibnamefont {Dahmen}}, \bibinfo
  {author} {\bibfnamefont {R.}~\bibnamefont {DeVore}}, \bibinfo {author}
  {\bibfnamefont {G.}~\bibnamefont {Petrova}},\ and\ \bibinfo {author}
  {\bibfnamefont {P.}~\bibnamefont {Wojtaszczyk}},\ }\bibfield  {title}
  {\bibinfo {title} {{Convergence Rates for Greedy Algorithms in Reduced Basis
  Methods}},\ }\href {https://doi.org/10.1137/100795772} {\bibfield  {journal}
  {\bibinfo  {journal} {SIAM J. Math. Anal.}\ }\textbf {\bibinfo {volume}
  {43}},\ \bibinfo {pages} {1457} (\bibinfo {year} {2011})}\BibitemShut
  {NoStop}%
\bibitem [{\citenamefont {den Nijs}\ and\ \citenamefont
  {Rommelse}(1989)}]{DenNijs1989}%
  \BibitemOpen
  \bibfield  {author} {\bibinfo {author} {\bibfnamefont {M.}~\bibnamefont {den
  Nijs}}\ and\ \bibinfo {author} {\bibfnamefont {K.}~\bibnamefont {Rommelse}},\
  }\bibfield  {title} {\bibinfo {title} {Preroughening transitions in crystal
  surfaces and valence-bond phases in quantum spin chains},\ }\href
  {https://doi.org/10.1103/PhysRevB.40.4709} {\bibfield  {journal} {\bibinfo
  {journal} {Phys. Rev. B}\ }\textbf {\bibinfo {volume} {40}},\ \bibinfo
  {pages} {4709} (\bibinfo {year} {1989})}\BibitemShut {NoStop}%
\bibitem [{\citenamefont {De~Chiara}\ \emph {et~al.}(2011)\citenamefont
  {De~Chiara}, \citenamefont {Lewenstein},\ and\ \citenamefont
  {Sanpera}}]{dechiaraBilinearbiquadraticSpin1Chain2011}%
  \BibitemOpen
  \bibfield  {author} {\bibinfo {author} {\bibfnamefont {G.}~\bibnamefont
  {De~Chiara}}, \bibinfo {author} {\bibfnamefont {M.}~\bibnamefont
  {Lewenstein}},\ and\ \bibinfo {author} {\bibfnamefont {A.}~\bibnamefont
  {Sanpera}},\ }\bibfield  {title} {\bibinfo {title} {{Bilinear-Biquadratic
  Spin-1 Chain Undergoing Quadratic Zeeman Effect}},\ }\href
  {https://doi.org/10.1103/PhysRevB.84.054451} {\bibfield  {journal} {\bibinfo
  {journal} {Phys. Rev. B}\ }\textbf {\bibinfo {volume} {84}},\ \bibinfo
  {pages} {054451} (\bibinfo {year} {2011})}\BibitemShut {NoStop}%
\bibitem [{\citenamefont {Rodr{\'i}guez}\ \emph {et~al.}(2011)\citenamefont
  {Rodr{\'i}guez}, \citenamefont {Arg{\"u}elles}, \citenamefont {Kolezhuk},
  \citenamefont {Santos},\ and\ \citenamefont
  {Vekua}}]{rodriguezFieldInducedPhaseTransitions2011}%
  \BibitemOpen
  \bibfield  {author} {\bibinfo {author} {\bibfnamefont {K.}~\bibnamefont
  {Rodr{\'i}guez}}, \bibinfo {author} {\bibfnamefont {A.}~\bibnamefont
  {Arg{\"u}elles}}, \bibinfo {author} {\bibfnamefont {A.~K.}\ \bibnamefont
  {Kolezhuk}}, \bibinfo {author} {\bibfnamefont {L.}~\bibnamefont {Santos}},\
  and\ \bibinfo {author} {\bibfnamefont {T.}~\bibnamefont {Vekua}},\ }\bibfield
   {title} {\bibinfo {title} {Field-{{Induced Phase Transitions}} of
  {{Repulsive Spin-1 Bosons}} in {{Optical Lattices}}},\ }\href
  {https://doi.org/10.1103/PhysRevLett.106.105302} {\bibfield  {journal}
  {\bibinfo  {journal} {Phys. Rev. Lett.}\ }\textbf {\bibinfo {volume} {106}},\
  \bibinfo {pages} {105302} (\bibinfo {year} {2011})}\BibitemShut {NoStop}%
\bibitem [{\citenamefont {F\'ath}\ and\ \citenamefont
  {S\'olyom}(1995)}]{Fath1995}%
  \BibitemOpen
  \bibfield  {author} {\bibinfo {author} {\bibfnamefont {G.}~\bibnamefont
  {F\'ath}}\ and\ \bibinfo {author} {\bibfnamefont {J.}~\bibnamefont
  {S\'olyom}},\ }\bibfield  {title} {\bibinfo {title} {Search for the
  nondimerized quantum nematic phase in the spin-1 chain},\ }\href
  {https://doi.org/10.1103/PhysRevB.51.3620} {\bibfield  {journal} {\bibinfo
  {journal} {Phys. Rev. B}\ }\textbf {\bibinfo {volume} {51}},\ \bibinfo
  {pages} {3620} (\bibinfo {year} {1995})}\BibitemShut {NoStop}%
\bibitem [{\citenamefont {Buchta}\ \emph {et~al.}(2005)\citenamefont {Buchta},
  \citenamefont {F\'ath}, \citenamefont {Legeza},\ and\ \citenamefont
  {S\'olyom}}]{Buchta2005}%
  \BibitemOpen
  \bibfield  {author} {\bibinfo {author} {\bibfnamefont {K.}~\bibnamefont
  {Buchta}}, \bibinfo {author} {\bibfnamefont {G.}~\bibnamefont {F\'ath}},
  \bibinfo {author} {\bibfnamefont {O.}~\bibnamefont {Legeza}},\ and\ \bibinfo
  {author} {\bibfnamefont {J.}~\bibnamefont {S\'olyom}},\ }\bibfield  {title}
  {\bibinfo {title} {Probable absence of a quadrupolar spin-nematic phase in
  the bilinear-biquadratic spin-1 chain},\ }\href
  {https://doi.org/10.1103/PhysRevB.72.054433} {\bibfield  {journal} {\bibinfo
  {journal} {Phys. Rev. B}\ }\textbf {\bibinfo {volume} {72}},\ \bibinfo
  {pages} {054433} (\bibinfo {year} {2005})}\BibitemShut {NoStop}%
\bibitem [{\citenamefont {Rizzi}\ \emph {et~al.}(2005)\citenamefont {Rizzi},
  \citenamefont {Rossini}, \citenamefont {De~Chiara}, \citenamefont
  {Montangero},\ and\ \citenamefont {Fazio}}]{Rizzi2005}%
  \BibitemOpen
  \bibfield  {author} {\bibinfo {author} {\bibfnamefont {M.}~\bibnamefont
  {Rizzi}}, \bibinfo {author} {\bibfnamefont {D.}~\bibnamefont {Rossini}},
  \bibinfo {author} {\bibfnamefont {G.}~\bibnamefont {De~Chiara}}, \bibinfo
  {author} {\bibfnamefont {S.}~\bibnamefont {Montangero}},\ and\ \bibinfo
  {author} {\bibfnamefont {R.}~\bibnamefont {Fazio}},\ }\bibfield  {title}
  {\bibinfo {title} {{Phase Diagram of Spin-1 Bosons on One-Dimensional
  Lattices}},\ }\href {https://doi.org/10.1103/PhysRevLett.95.240404}
  {\bibfield  {journal} {\bibinfo  {journal} {Phys. Rev. Lett.}\ }\textbf
  {\bibinfo {volume} {95}},\ \bibinfo {pages} {240404} (\bibinfo {year}
  {2005})}\BibitemShut {NoStop}%
\bibitem [{\citenamefont {L\"auchli}\ \emph {et~al.}(2006)\citenamefont
  {L\"auchli}, \citenamefont {Schmid},\ and\ \citenamefont
  {Trebst}}]{Laeuchli2006}%
  \BibitemOpen
  \bibfield  {author} {\bibinfo {author} {\bibfnamefont {A.}~\bibnamefont
  {L\"auchli}}, \bibinfo {author} {\bibfnamefont {G.}~\bibnamefont {Schmid}},\
  and\ \bibinfo {author} {\bibfnamefont {S.}~\bibnamefont {Trebst}},\
  }\bibfield  {title} {\bibinfo {title} {Spin nematics correlations in
  bilinear-biquadratic ${S}=1$ spin chains},\ }\href
  {https://doi.org/10.1103/PhysRevB.74.144426} {\bibfield  {journal} {\bibinfo
  {journal} {Phys. Rev. B}\ }\textbf {\bibinfo {volume} {74}},\ \bibinfo
  {pages} {144426} (\bibinfo {year} {2006})}\BibitemShut {NoStop}%
\bibitem [{\citenamefont {Hu}\ \emph {et~al.}(2014)\citenamefont {Hu},
  \citenamefont {Turner}, \citenamefont {Penc},\ and\ \citenamefont
  {Pollmann}}]{Hu2014}%
  \BibitemOpen
  \bibfield  {author} {\bibinfo {author} {\bibfnamefont {S.}~\bibnamefont
  {Hu}}, \bibinfo {author} {\bibfnamefont {A.~M.}\ \bibnamefont {Turner}},
  \bibinfo {author} {\bibfnamefont {K.}~\bibnamefont {Penc}},\ and\ \bibinfo
  {author} {\bibfnamefont {F.}~\bibnamefont {Pollmann}},\ }\bibfield  {title}
  {\bibinfo {title} {{Berry-Phase-Induced Dimerization in One-Dimensional
  Quadrupolar Systems}},\ }\href
  {https://doi.org/10.1103/PhysRevLett.113.027202} {\bibfield  {journal}
  {\bibinfo  {journal} {Phys. Rev. Lett.}\ }\textbf {\bibinfo {volume} {113}},\
  \bibinfo {pages} {027202} (\bibinfo {year} {2014})}\BibitemShut {NoStop}%
\bibitem [{\citenamefont {Weyrauch}\ and\ \citenamefont
  {Rakov}(2018)}]{Weyrauch2018}%
  \BibitemOpen
  \bibfield  {author} {\bibinfo {author} {\bibfnamefont {M.}~\bibnamefont
  {Weyrauch}}\ and\ \bibinfo {author} {\bibfnamefont {M.~V.}\ \bibnamefont
  {Rakov}},\ }\bibfield  {title} {\bibinfo {title} {Dimerization in the
  anisotropic bilinear-biquadratic {Heisenberg} model},\ }\href
  {https://doi.org/10.1088/1742-6596/1041/1/012004} {\bibfield  {journal}
  {\bibinfo  {journal} {J. Phys. Conf. Ser.}\ }\textbf {\bibinfo {volume}
  {1041}},\ \bibinfo {pages} {012004} (\bibinfo {year} {2018})}\BibitemShut
  {NoStop}%
\bibitem [{\citenamefont {Eftang}\ \emph {et~al.}(2010)\citenamefont {Eftang},
  \citenamefont {Patera},\ and\ \citenamefont {R\o{}nquist}}]{Eftang2010}%
  \BibitemOpen
  \bibfield  {author} {\bibinfo {author} {\bibfnamefont {J.~L.}\ \bibnamefont
  {Eftang}}, \bibinfo {author} {\bibfnamefont {A.~T.}\ \bibnamefont {Patera}},\
  and\ \bibinfo {author} {\bibfnamefont {E.~M.}\ \bibnamefont {R\o{}nquist}},\
  }\bibfield  {title} {\bibinfo {title} {{An "$hp$" Certified Reduced Basis
  Method for Parametrized Elliptic Partial Differential Equations}},\ }\href
  {https://doi.org/10.1137/090780122} {\bibfield  {journal} {\bibinfo
  {journal} {SIAM J. Sci. Comp.}\ }\textbf {\bibinfo {volume} {32}},\ \bibinfo
  {pages} {3170} (\bibinfo {year} {2010})}\BibitemShut {NoStop}%
\bibitem [{\citenamefont {Eftang}\ \emph {et~al.}(2011)\citenamefont {Eftang},
  \citenamefont {Knezevic},\ and\ \citenamefont {Patera}}]{Eftang2011}%
  \BibitemOpen
  \bibfield  {author} {\bibinfo {author} {\bibfnamefont {J.~L.}\ \bibnamefont
  {Eftang}}, \bibinfo {author} {\bibfnamefont {D.~J.}\ \bibnamefont
  {Knezevic}},\ and\ \bibinfo {author} {\bibfnamefont {A.~T.}\ \bibnamefont
  {Patera}},\ }\bibfield  {title} {\bibinfo {title} {An $hp$ certified reduced
  basis method for parametrized parabolic partial differential equations},\
  }\href {https://doi.org/10.1080/13873954.2011.547670} {\bibfield  {journal}
  {\bibinfo  {journal} {Math. Comp. Model. Dyn. Syst.}\ }\textbf {\bibinfo
  {volume} {17}},\ \bibinfo {pages} {395} (\bibinfo {year} {2011})}\BibitemShut
  {NoStop}%
\bibitem [{\citenamefont {Eftang}\ and\ \citenamefont
  {Stamm}(2012)}]{Eftang2012}%
  \BibitemOpen
  \bibfield  {author} {\bibinfo {author} {\bibfnamefont {J.~L.}\ \bibnamefont
  {Eftang}}\ and\ \bibinfo {author} {\bibfnamefont {B.}~\bibnamefont {Stamm}},\
  }\bibfield  {title} {\bibinfo {title} {Parameter multi-domain ‘hp’
  empirical interpolation},\ }\href
  {https://doi.org/https://doi.org/10.1002/nme.3327} {\bibfield  {journal}
  {\bibinfo  {journal} {Int. J. Numer. Methods Eng.}\ }\textbf {\bibinfo
  {volume} {90}},\ \bibinfo {pages} {412} (\bibinfo {year} {2012})}\BibitemShut
  {NoStop}%
\bibitem [{\citenamefont {Zanardi}\ and\ \citenamefont
  {Paunkovi\ifmmode~\acute{c}\else \'{c}\fi{}}(2006)}]{Zanardi2006}%
  \BibitemOpen
  \bibfield  {author} {\bibinfo {author} {\bibfnamefont {P.}~\bibnamefont
  {Zanardi}}\ and\ \bibinfo {author} {\bibfnamefont {N.}~\bibnamefont
  {Paunkovi\ifmmode~\acute{c}\else \'{c}\fi{}}},\ }\bibfield  {title} {\bibinfo
  {title} {Ground state overlap and quantum phase transitions},\ }\href
  {https://doi.org/10.1103/PhysRevE.74.031123} {\bibfield  {journal} {\bibinfo
  {journal} {Phys. Rev. E}\ }\textbf {\bibinfo {volume} {74}},\ \bibinfo
  {pages} {031123} (\bibinfo {year} {2006})}\BibitemShut {NoStop}%
\bibitem [{\citenamefont {You}\ \emph {et~al.}(2007)\citenamefont {You},
  \citenamefont {Li},\ and\ \citenamefont {Gu}}]{Gu2007}%
  \BibitemOpen
  \bibfield  {author} {\bibinfo {author} {\bibfnamefont {W.-L.}\ \bibnamefont
  {You}}, \bibinfo {author} {\bibfnamefont {Y.-W.}\ \bibnamefont {Li}},\ and\
  \bibinfo {author} {\bibfnamefont {S.-J.}\ \bibnamefont {Gu}},\ }\bibfield
  {title} {\bibinfo {title} {Fidelity, dynamic structure factor, and
  susceptibility in critical phenomena},\ }\href
  {https://doi.org/10.1103/PhysRevE.76.022101} {\bibfield  {journal} {\bibinfo
  {journal} {Phys. Rev. E}\ }\textbf {\bibinfo {volume} {76}},\ \bibinfo
  {pages} {022101} (\bibinfo {year} {2007})}\BibitemShut {NoStop}%
\bibitem [{\citenamefont {Gu}(2010)}]{Gu2010}%
  \BibitemOpen
  \bibfield  {author} {\bibinfo {author} {\bibfnamefont {S.-J.}\ \bibnamefont
  {Gu}},\ }\bibfield  {title} {\bibinfo {title} {Fidelity approach to quantum
  phase transitions},\ }\href {https://doi.org/10.1142/S0217979210056335}
  {\bibfield  {journal} {\bibinfo  {journal} {Int. J. Mod. Phys. B}\ }\textbf
  {\bibinfo {volume} {24}},\ \bibinfo {pages} {4371} (\bibinfo {year}
  {2010})}\BibitemShut {NoStop}%
\bibitem [{\citenamefont {Eisert}\ \emph {et~al.}(2010)\citenamefont {Eisert},
  \citenamefont {Cramer},\ and\ \citenamefont {Plenio}}]{Eisert2010}%
  \BibitemOpen
  \bibfield  {author} {\bibinfo {author} {\bibfnamefont {J.}~\bibnamefont
  {Eisert}}, \bibinfo {author} {\bibfnamefont {M.}~\bibnamefont {Cramer}},\
  and\ \bibinfo {author} {\bibfnamefont {M.~B.}\ \bibnamefont {Plenio}},\
  }\bibfield  {title} {\bibinfo {title} {Colloquium: Area laws for the
  entanglement entropy},\ }\href {https://doi.org/10.1103/RevModPhys.82.277}
  {\bibfield  {journal} {\bibinfo  {journal} {Rev. Mod. Phys.}\ }\textbf
  {\bibinfo {volume} {82}},\ \bibinfo {pages} {277} (\bibinfo {year}
  {2010})}\BibitemShut {NoStop}%
\bibitem [{\citenamefont {Zeng}\ \emph {et~al.}(2019)\citenamefont {Zeng},
  \citenamefont {Chen}, \citenamefont {Zhou},\ and\ \citenamefont
  {Wen}}]{Zeng2019}%
  \BibitemOpen
  \bibfield  {author} {\bibinfo {author} {\bibfnamefont {B.}~\bibnamefont
  {Zeng}}, \bibinfo {author} {\bibfnamefont {X.}~\bibnamefont {Chen}}, \bibinfo
  {author} {\bibfnamefont {D.-L.}\ \bibnamefont {Zhou}},\ and\ \bibinfo
  {author} {\bibfnamefont {X.-G.}\ \bibnamefont {Wen}},\ }\href
  {https://doi.org/10.1007/978-1-4939-9084-9} {\emph {\bibinfo {title} {Quantum
  Information Meets Quantum Matter}}},\ Quantum {{Science}} and {{Technology}}\
  (\bibinfo  {publisher} {{Springer}},\ \bibinfo {address} {{New York, NY}},\
  \bibinfo {year} {2019})\BibitemShut {NoStop}%
\bibitem [{\citenamefont {Song}\ \emph {et~al.}(2010)\citenamefont {Song},
  \citenamefont {Rachel},\ and\ \citenamefont {Le~Hur}}]{Song2010}%
  \BibitemOpen
  \bibfield  {author} {\bibinfo {author} {\bibfnamefont {H.~F.}\ \bibnamefont
  {Song}}, \bibinfo {author} {\bibfnamefont {S.}~\bibnamefont {Rachel}},\ and\
  \bibinfo {author} {\bibfnamefont {K.}~\bibnamefont {Le~Hur}},\ }\bibfield
  {title} {\bibinfo {title} {General relation between entanglement and
  fluctuations in one dimension},\ }\href
  {https://doi.org/10.1103/PhysRevB.82.012405} {\bibfield  {journal} {\bibinfo
  {journal} {Phys. Rev. B}\ }\textbf {\bibinfo {volume} {82}},\ \bibinfo
  {pages} {012405} (\bibinfo {year} {2010})}\BibitemShut {NoStop}%
\bibitem [{\citenamefont {Song}\ \emph {et~al.}(2012)\citenamefont {Song},
  \citenamefont {Rachel}, \citenamefont {Flindt}, \citenamefont {Klich},
  \citenamefont {Laflorencie},\ and\ \citenamefont {Le~Hur}}]{Song2012}%
  \BibitemOpen
  \bibfield  {author} {\bibinfo {author} {\bibfnamefont {H.~F.}\ \bibnamefont
  {Song}}, \bibinfo {author} {\bibfnamefont {S.}~\bibnamefont {Rachel}},
  \bibinfo {author} {\bibfnamefont {C.}~\bibnamefont {Flindt}}, \bibinfo
  {author} {\bibfnamefont {I.}~\bibnamefont {Klich}}, \bibinfo {author}
  {\bibfnamefont {N.}~\bibnamefont {Laflorencie}},\ and\ \bibinfo {author}
  {\bibfnamefont {K.}~\bibnamefont {Le~Hur}},\ }\bibfield  {title} {\bibinfo
  {title} {Bipartite fluctuations as a probe of many-body entanglement},\
  }\href {https://doi.org/10.1103/PhysRevB.85.035409} {\bibfield  {journal}
  {\bibinfo  {journal} {Phys. Rev. B}\ }\textbf {\bibinfo {volume} {85}},\
  \bibinfo {pages} {035409} (\bibinfo {year} {2012})}\BibitemShut {NoStop}%
\bibitem [{\citenamefont {Rachel}\ \emph {et~al.}(2012)\citenamefont {Rachel},
  \citenamefont {Laflorencie}, \citenamefont {Song},\ and\ \citenamefont
  {Le~Hur}}]{Rachel2012}%
  \BibitemOpen
  \bibfield  {author} {\bibinfo {author} {\bibfnamefont {S.}~\bibnamefont
  {Rachel}}, \bibinfo {author} {\bibfnamefont {N.}~\bibnamefont {Laflorencie}},
  \bibinfo {author} {\bibfnamefont {H.~F.}\ \bibnamefont {Song}},\ and\
  \bibinfo {author} {\bibfnamefont {K.}~\bibnamefont {Le~Hur}},\ }\bibfield
  {title} {\bibinfo {title} {{Detecting Quantum Critical Points Using Bipartite
  Fluctuations}},\ }\href {https://doi.org/10.1103/PhysRevLett.108.116401}
  {\bibfield  {journal} {\bibinfo  {journal} {Phys. Rev. Lett.}\ }\textbf
  {\bibinfo {volume} {108}},\ \bibinfo {pages} {116401} (\bibinfo {year}
  {2012})}\BibitemShut {NoStop}%
\bibitem [{\citenamefont {Contessi}\ \emph {et~al.}(2023)\citenamefont
  {Contessi}, \citenamefont {Recati},\ and\ \citenamefont
  {Rizzi}}]{Contessi2023}%
  \BibitemOpen
  \bibfield  {author} {\bibinfo {author} {\bibfnamefont {D.}~\bibnamefont
  {Contessi}}, \bibinfo {author} {\bibfnamefont {A.}~\bibnamefont {Recati}},\
  and\ \bibinfo {author} {\bibfnamefont {M.}~\bibnamefont {Rizzi}},\ }\bibfield
   {title} {\bibinfo {title} {Phase diagram detection via gaussian fitting of
  number probability distribution},\ }\href
  {https://doi.org/10.1103/PhysRevB.107.L121403} {\bibfield  {journal}
  {\bibinfo  {journal} {Phys. Rev. B}\ }\textbf {\bibinfo {volume} {107}},\
  \bibinfo {pages} {L121403} (\bibinfo {year} {2023})}\BibitemShut {NoStop}%
\end{thebibliography}%

\end{document}